\documentclass[aps,prx,superscriptaddress,showpacs,twocolumn]{revtex4-2}
\usepackage[left=1.9cm,right=1.9cm,bottom=1.9cm,top=1.9cm,ignoreall]{geometry}
\usepackage[utf8]{inputenc}
\usepackage{subfigure}
\usepackage{graphicx}
\usepackage[skip=0pt, indent=20pt]{parskip}
\usepackage{amsmath}
\usepackage{amssymb}
\usepackage{xspace}
\usepackage{bm}
\usepackage{color}
\usepackage[squaren,Gray]{SIunits}
\usepackage[hyperindex=true]{hyperref}
\hypersetup{linktocpage,colorlinks=true,citecolor=blue,linkcolor=blue}
\usepackage{empheq}
\usepackage{braket}
\usepackage{physics}

\usepackage{microtype} % pour éviter hyphenations trop courtes
\begin{document}
	
\title{Hybrid III/V-Silicon quantum photonic device generating \\broadband entangled photon pairs}

\author{J. Schuhmann}
\thanks{These authors contributed equally to this work.}
\affiliation{Laboratoire Matériaux et Phénomènes Quantiques, Université Paris Cité, 75013 Paris, France}
\affiliation{Centre de Nanosciences et Nanotechnologies, Université Paris-Saclay, 91120 Palaiseau, France}
\affiliation{STMicroelectronics, Technology $\And$ Design Platform, 38920 Crolles, France}

\author{L. Lazzari}
\thanks{These authors contributed equally to this work.}
\affiliation{Laboratoire Matériaux et Phénomènes Quantiques, Université Paris Cité, 75013 Paris, France}
\affiliation{Centre de Nanosciences et Nanotechnologies, Université Paris-Saclay, 91120 Palaiseau, France}
\affiliation{STMicroelectronics, Technology $\And$ Design Platform, 38920 Crolles, France}

\author{M. Morassi}
\affiliation{Centre de Nanosciences et Nanotechnologies, Université Paris-Saclay, 91120 Palaiseau, France}

\author{A. Lemaître}
\affiliation{Centre de Nanosciences et Nanotechnologies, Université Paris-Saclay, 91120 Palaiseau, France}

\author{I. Sagnes} 
\affiliation{Centre de Nanosciences et Nanotechnologies, Université Paris-Saclay, 91120 Palaiseau, France}

\author{G.~Beaudoin}
\affiliation{Centre de Nanosciences et Nanotechnologies, Université Paris-Saclay, 91120 Palaiseau, France}

\author{M.I. Amanti}
\affiliation{Laboratoire Matériaux et Phénomènes Quantiques, Université Paris Cité, 75013 Paris, France}

\author{F. Boeuf}
\affiliation{STMicroelectronics, Technology $\And$ Design Platform, 38920 Crolles, France}

\author{F. Raineri}
\affiliation{Centre de Nanosciences et Nanotechnologies, Université Paris-Saclay, 91120 Palaiseau, France}
\affiliation{Université Côte d’Azur, Institut de Physique de Nice, CNRS-UMR 7010, Sophia Antipolis, France}

\author{F. Baboux}
\thanks{Corresponding author: \textcolor{blue}{florent.baboux@u-paris.fr}}
\affiliation{Laboratoire Matériaux et Phénomènes Quantiques, Université Paris Cité, 75013 Paris, France}

\author{S. Ducci}
\affiliation{Laboratoire Matériaux et Phénomènes Quantiques, Université Paris Cité, 75013 Paris, France}

\makeatletter
\def\Dated@name{} % to remove "dated:"
\makeatother

\begin{abstract}

The demand for integrated photonic chips combining the generation and manipulation of quantum states of light is steadily increasing, driven by the need for compact and scalable platforms for quantum information technologies. While photonic circuits with diverse functionalities are being developed in different single material platforms, it has become crucial to realize hybrid photonic circuits that harness the advantages of multiple materials while mitigating their respective weaknesses, resulting in enhanced capabilities. Here, we demonstrate a hybrid III-V/Silicon quantum photonic device combining the strong second-order nonlinearity and direct bandgap of the III-V semiconductor platform with the high maturity and CMOS compatibility of the silicon photonic platform. Our device embeds the spontaneous parametric down-conversion (SPDC) of photon pairs into an AlGaAs source and their vertical routing to an adhesively-bonded silicon-on-insulator circuitry, within an evanescent coupling scheme managing both polarization states. This enables the on-chip generation of broadband ($>\!$  40 nm) telecom photons by type 0 and type 2 SPDC from the hybrid device, at room temperature and with internal pair generation rates exceeding $10^5$ $s^{-1}$ for both types, while the pump beam is strongly rejected. Two-photon interference with 92\% visibility (and up to 99\% upon 5 nm spectral filtering) proves the high energy-time entanglement quality of the produced quantum state, thereby enabling a wide range of quantum information applications on-chip, within an hybrid architecture compliant with electrical pumping and merging the assets of two mature and highly complementary platforms in view of out-of-the-lab deployment of quantum technologies.

\end{abstract}

\maketitle

\section*{Introduction}

Quantum photonics is emerging as a pivotal actor in the development of quantum technologies, driving transformative changes in various fields such as communication, computing and simulation tasks \cite{Flamini19}. 
Indeed, the utilization of photons as information carriers presents an ideal solution due to their inherent robustness to noise, high propagation speed and large variety of degrees of freedom to encode information. In addition, photons can be efficiently generated and waveguided into miniaturized chip-integrated structures, which provides an increased compacity and portability, but also a gain of stability allowing to scale up the complexity of the operations carried out \cite{Wang20,Pelucchi22}. 

The development of chip-scale quantum photonic circuits has thus become essential to move from laboratory experiments to large-scale real-world implementations. The diverse requirements of each specific application led to the exploration of various photonics platforms in the last decades, including silicon-based materials \cite{Silverstone16}, III-V semiconductors \cite{Dietrich16}, dielectrics as lithium niobate \cite{Alibart16} as well as various other emerging materials \cite{Elshaari20}. However, the implementation of quantum information tasks requires competing functionalities that challenge the realization of integrated quantum photonic circuits combining the generation, manipulation and detection of quantum states of light from a single monolithic material platform. Therefore, significant efforts are currently focused on combining different platforms to create hybrid photonic circuits leveraging the strengths of different materials while avoiding their respective weaknesses, so as to reach enhanced capabilities \cite{Kim20,Elshaari20}.

Among the various possible combinations, the hybridization of III-V compounds (such as GaAs and InP) with silicon-based materials (silicon-on-insulator, silica-on-silicon, silicon nitride) holds great promise. Indeed, while the highly mature silicon-based platform allows realizing high-quality and low-loss circuits with access to a wide variety of optical components \cite{Silverstone16,Review_Siew}, it suffers from an important weakness, namely its indirect bandgap that complicates the implementation of efficient light sources. By contrast, III-V compounds like GaAs or InP are, thanks to their direct bandgap, ideal materials to realize low-cost and efficient lasers, amplifiers and photodetectors; in addition, their high electro-optic effect allows realizing fast modulators, and they provide for quantum photonics the capability to implement both high-quality single-photon emitters and parametric sources \cite{Dietrich16,Baboux23}. The silicon and III-V platforms are thus highly complementary, which motivated numerous technological efforts to combine them in recent years.

In classical photonics, the realization of hybrid \mbox{III-V/silicon} lasers is highly sought-after due the difficulty of realizing purely Si-based lasers \cite{Liang10}. In quantum photonics, the incorporation of III-V single-photon emitters into silicon-based circuits has been achieved in recent years, allowing to integrate the emission, routing and manipulation of quantum states of light in hybrid circuits \cite{Kim20,Elshaari20}. In particular, InAs quantum dots have been integrated with silicon nitride waveguides \cite{Davanco17} and silica-on-silicon microdisks \cite{Yue18} by bonding techniques, and with silicon-on-insulator waveguides \cite{Osada19} by transfer printing; on the other hand, the deterministic transfer of single InAs/InP quantum dots embedded in InP nanowires in silicon nitride~\cite{Zadeh16} and silicon-on-insulator \cite{Kim17} circuits has been carried out by pick-and-place techniques. 

Hence, significant efforts have been focused on the integration of III-V single-photon emitters into silicon-based circuits; by contrast, the integration of \mbox{III-V} parametric sources has been scarcely explored so far. Yet, it would bring along unique and complementary advantages such as room temperature operation, a high fabrication reproducibility enabling the realization of a large number of identical sources on the same chip, and a high quality and versatility of the produced quantum states -- either heralded single photons \cite{Aharonovich16,Belhassen18}, entangled photon pairs \cite{Chen18,Wang21} or squeezed states \cite{Andersen16,Brodutch18}.
Recently, a packaged hybrid InP/SiN device combining an electrically pumped laser and SFWM generation of photon pair in a SiN microring has been reported \cite{kues23}, featuring a $8.2 \times 10^3 s^{-1}$ internal pair generation rate and $\sim 8$ nm bandwidth; a cascade of microring resonators is used to suppress the pump beam, which lies at the same frequency than the down-converted photons.
Recently, an hybrid chip consisting of an electrically pumped InP gain section followed by a SiN microring generating photon pairs by SFWM has been reported~\cite{kues23}, featuring a $8.2 \! \times \! 10^3$ s$^{-1}$ internal pair generation rate and $\sim \! 8$ nm bandwidth; a cascade of microring resonators is used to suppress the pump beam, which lies at the same frequency than the down-converted photons.
We demonstrate here an alternative approach, by directly integrating an AlGaAs SPDC source on top of a silicon-on-insulator waveguide structure, to realize a hybrid quantum photonic device including photon pair generation and routing at room temperature and telecom wavelength. 
Photon pairs are generated in the AlGaAs waveguide and routed through evanescent coupling to a silicon-on-insulator waveguide lying underneath. The use of SPDC, where the pump and down-converted photons are well separated in frequency, allows an intrinsic filtering of the pump beam in a simple structure design. Thanks to the polarization versatility of our coupling scheme, our hybrid III-V/Silicon device allows the operation of both type-0 and type-2 SPDC processes, leading to the emission of broadband ($> \!$ 40 nm) telecom photon pairs with an internal pair generation rate exceeding $10^5 \!$ s$^{-1}$ for both types. Two-photon interference with 92\% visibility (up to 99\% upon 5 nm spectral filtering) demonstrates the high energy-time entanglement of the produced quantum state, opening the way to a wide variety of quantum information applications on-chip.

\begin{figure*}[t]
    \includegraphics[width= \textwidth]{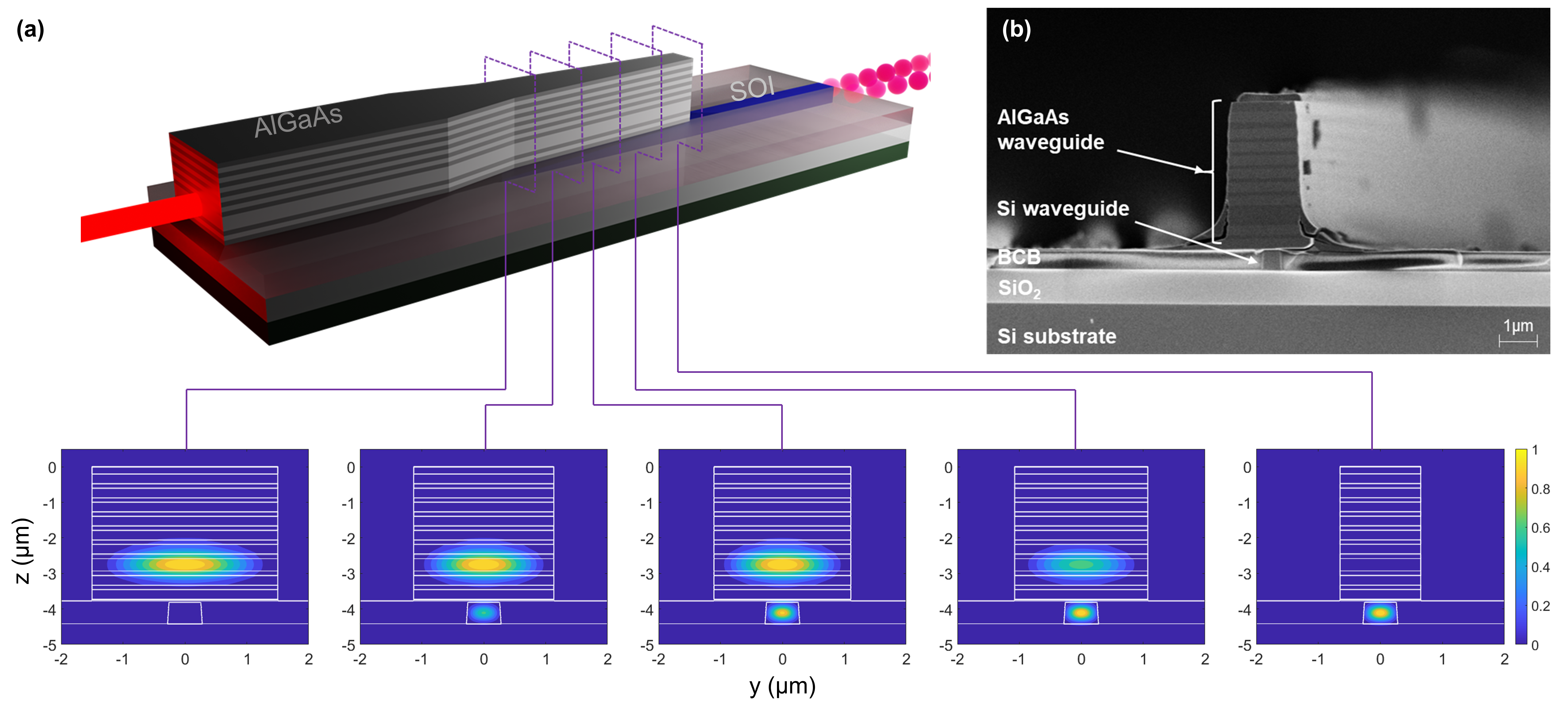}
    \caption{(a) Sketch of the hybrid III-V/Silicon structure (not in scale) with numerical simulations of the mode coupling profile (intensity) at different positions along the structure, for TM polarization at 1550 nm wavelength (similar results are obtained for TE polarization). (b) Transverse SEM image of the fabricated hybrid structure, in the coupling region.}
     \label{fig:coupling_sections}
\end{figure*}

\section*{Working principle and design}

The working principle of our hybrid device is sketched in Fig. \ref{fig:coupling_sections}a: photon pairs, generated upon optical pumping in an AlGaAs waveguide (shown in shades of gray) are transferred to the SOI circuitry (blue) by evanescent coupling, preserving the properties of the produced quantum state. We describe below the various components of this hybrid device successively: the AlGaAs parametric source, the SOI circuitry and the coupling scheme.

The AlGaAs source is a nano-fabricated waveguide generating bi-photon states via spontaneous parametric down-conversion (SPDC): a pump beam at frequency $\omega_p$ is coupled into the waveguide and undergoes a down-conversion process producing pairs of photons at frequencies $\omega_s$ and $\omega_i$, such that $\omega_p=\omega_s+\omega_i$ for energy conservation. Bragg mirrors, obtained by tuning the aluminum concentration of the different AlGaAs layers, provide both a photonic bandgap confinement for the pump in the near-infrared range (NIR, $\sim775$ nm) and total internal confinement for the produced photons in the telecom range ($\sim1550$ nm) \cite{BRW_Original,Review_Feli}. Therefore, the pump and SPDC modes are characterized by different dispersion curves, allowing the phase-matching condition to be satisfied in the spectral range of interest. The source is designed to optimize the nonlinear conversion efficiency for two phase-matching processes: type 0 SPDC, where a TM polarized pump beam generates TM polarized photons, and type 2, where a TE pump beam generates orthogonally polarized photons. The structure, grown by molecular beam epitaxy on a GaAs substrate, features a 364 nm-thick core made of Al$_{0.45}$Ga$_{0.55}$As sandwiched between the two Bragg mirrors. The latter consist of alternating 116 nm-thick Al$_{0.25}$Ga$_{0.75}$As layers and 280 nm-thick Al$_{0.8}$Ga$_{0.2}$As layers, these parameters being chosen to maximize the nonlinear integral overlap of the parametric process.
The mirrors are asymmetric: 2 pairs of layers are used on the side that will be subsequently bonded to the SOI structure, and 6 pairs on the other side; this ensures a good mode confinement -- no relevant reduction is observed compared to a symmetric \mbox{structure --,} while resulting more suited to hybridization, as detailed below.

The bi-photon state generated via SPDC can be written as:
\begin{equation}
    \ket\psi=\iint \text{d}\omega_s\text{d}\omega_i \, C(\omega_s,\omega_i) \,  \hat{a}_s^\dag(\omega_s)\hat{a}_i^\dag(\omega_i)\ket0, \label{eq:stateJSA}
\end{equation}
where $\hat{a}_x^\dag(\omega_x)$ is the operator creating a photon in the mode $x$ with frequency $\omega_x$ and $\ket0$ is the vacuum state. $C(\omega_s,\omega_i)$ is the joint spectral amplitude (JSA), whose modulus squared -- the joint spectral intensity (JSI) -- gives the probability that the produced state is composed of a signal photon at frequency $\omega_s$ and an idler photon at frequency $\omega_i$. 
Neglecting group velocity dispersion and in the limit of narrow pump bandwidth, the JSA can be factorized \cite{Ansari18b,Gianani20} as $C(\omega_s,\omega_i)=\alpha_{\rm p}(\omega_s+\omega_i) \,  \Phi_{\rm PM}(\omega_s,\omega_i)$, where $\alpha_{\rm p}$ is the pump spectrum envelope and $\Phi_{\rm PM}$ reflects the phase-matching condition. In the following experiments, a narrowand CW laser is used (linewidth 100 kHz) so that $\alpha_{\rm p}$  can be considered as a Dirac delta and the JSA only depends of the frequency difference $\omega_s-\omega_i$, in excellent approximation.
Such CW pumping makes the photons generated through SPDC to be naturally energy-time and frequency-bin entangled~\cite{Maltese20}. Furthermore, photons produced via a type 2  process are also polarization entangled directly at the chip output, thanks to the very low modal birefringence of the waveguide \cite{Appas21}.

Now turning to the SOI part of the circuit, it consists in a 780 $\mu$m-thick Si substrate, a 1 $\mu$m-thick SiO$_2$ buried oxyde layer and a 610 nm-thick Si top layer. Silicon is transparent to wavelengths $\gtrsim 1$ $\mu$m, so that telecom optical modes can be confined and propagated in waveguides, while NIR radiation, corresponding to the SPDC pump for the AlGaAs source, is strongly absorbed \cite{Silicon_absCoeff}. The bottom width of the waveguides is set to 560 nm, resulting in a low modal birefringence ($<$ 1\%). Due to the large top layer thickness, the slope of the waveguide flanks represents an additional  and useful parameter for the fine control of the effective mode index, facilitating the mode coupling for both polarizations.

The hybrid structure is designed to achieve the optical mode transfer from the AlGaAs waveguide (5.5 µm wide, 1.6 mm long) to the silicon one (2.5 µm wide, 1~mm long), which are superimposed along the longitudinal direction corresponding to the mode propagation (Fig.~\ref{fig:coupling_sections}). 
An adiabatic coupling scheme has been chosen for its high coupling efficiency with minimized mode conversion to higher-order modes or radiation modes, together with minimized oscillations of the transmission as a function of the coupling length \cite{Sun09,Fu14}. This ensures enhanced fabrication tolerances compared e.g. to directional coupling, which in addition would require extended effective mode index matching between the waveguides (Appendix B).

We designed a linear taper in the AlGaAs waveguide to gradually and adiabatically transfer the optical power to the SOI waveguide, as shown in the mode profile simulations of Fig. \ref{fig:coupling_sections}a (bottom row). 
The taper is designed to achieve the transfer for both TE and TM modes. After the SPDC generation region, the AlGaAs waveguide width is first tapered down from 5.5 $\mu$m to 2.2 $\mu$m; in the coupling region, where the two waveguides are superimposed vertically, the width is then linearly decreased from 2.2 $\mu$m to 1.4 $\mu$m along 600 $\mu$m of coupling length. The silicon waveguide flanks present a slope of around 2.5°, corresponding to the etching angle produced by our dry etching process detailed below. Numerical simulations, performed with a commercial FDE (Finite-Difference Eigenmode) solver, predict peak transmissions higher than 70\% for both polarizations, over a broad bandwidth centered on the telecom C band (see Appendix C).

\section*{Sample fabrication and characterization}

Various strategies have been developed for facing the challenges of integrating III-V and SOI platforms. Direct (or molecular) bonding \cite{MolecularBonding} and metallic bonding \cite{Tang2018InvestigationAO} have emerged as powerful approaches; however, they are very sensitive to surface micro-roughness and require high processing temperatures. Furthermore, the metallic layer, being opaque to the telecom radiation, is not functional to the mode coupling. Alternative promising options are the direct epitaxy re-growth \cite{DirectRegrowth} and the micro-transfer printing \cite{MicroTransfertPrint}: the first technique demands the compensation of the lattice mismatch, often obtained by the use of a few hundreds of nanometers thick buffer layer that may affect the evanescent coupling; the second one suffers from a limited alignment accuracy, around 1 $\mu$m. These constraints are overcome by the adhesive bonding method that we here use \cite{Crosnier17}, consisting of bonding the two materials using a low-loss polymer (benzocyclobuten, BCB), resulting highly robust against surface roughness and requiring lower temperatures compared to the other techniques; in addition, the polymer thickness can be finely controlled and set to few tens of nanometers, in accordance with the coupling necessities.

The device fabrication involves several steps. First, the SOI circuit is prepared: the waveguides are patterned via electron beam lithography and then etched via inductive coupled plasma-deep reactive ion etching (ICP-DRIE); this technique provides fine control of the slope of the waveguide flanks, as required by the coupling design. Alignment marks are also etched on the structure during this step. The surface is treated with O$_2$ plasma and an adhesion promoter is applied. The BCB is then spin-coated onto the SOI platform to reach a 40 nm thickness. Meanwhile, a 20 nm film of SiO$_2$ is deposited onto the bonding side of the AlGaAs epitaxial structure via atomic layer deposition (ALD), followed by the application of the same adhesion promoter. The two materials are then superimposed one to the other, facing the treated sides. By heating up the system and applying a compressive force, the polymerization of the BCB turns it from liquid phase into a rigid bond between the two structures. The GaAs substrate is then selectively removed using a citric acid solution, and AlGaAs waveguides are patterned via e-beam lithography and etched by ICP-RIE, in alignment with the silicon waveguides beneath. An SEM image of the section belonging to the coupling region -- where the two waveguides overlap -- of the resulting device is shown in Fig. \ref{fig:coupling_sections}b.

The coupling design and the fabrication process are first validated by measuring the transmitted optical power in the telecom range, using a linearly polarized CW laser.  Considering the estimated injection and collection efficiency, waveguide facet reflectivities (numerically simulated with a commercial FDTD solver) and optical losses, we can retrieve the actual mode coupling efficiency (Appendix C). The transmission spectra, centered at different wavelengths, reach $\simeq 60\%$  and present a FWHM $> \, 70 $ nm for both TE and TM polarizations, thus validating our coupling design.

\begin{figure*}[t]
    \centering
    \includegraphics[width=0.9\textwidth]{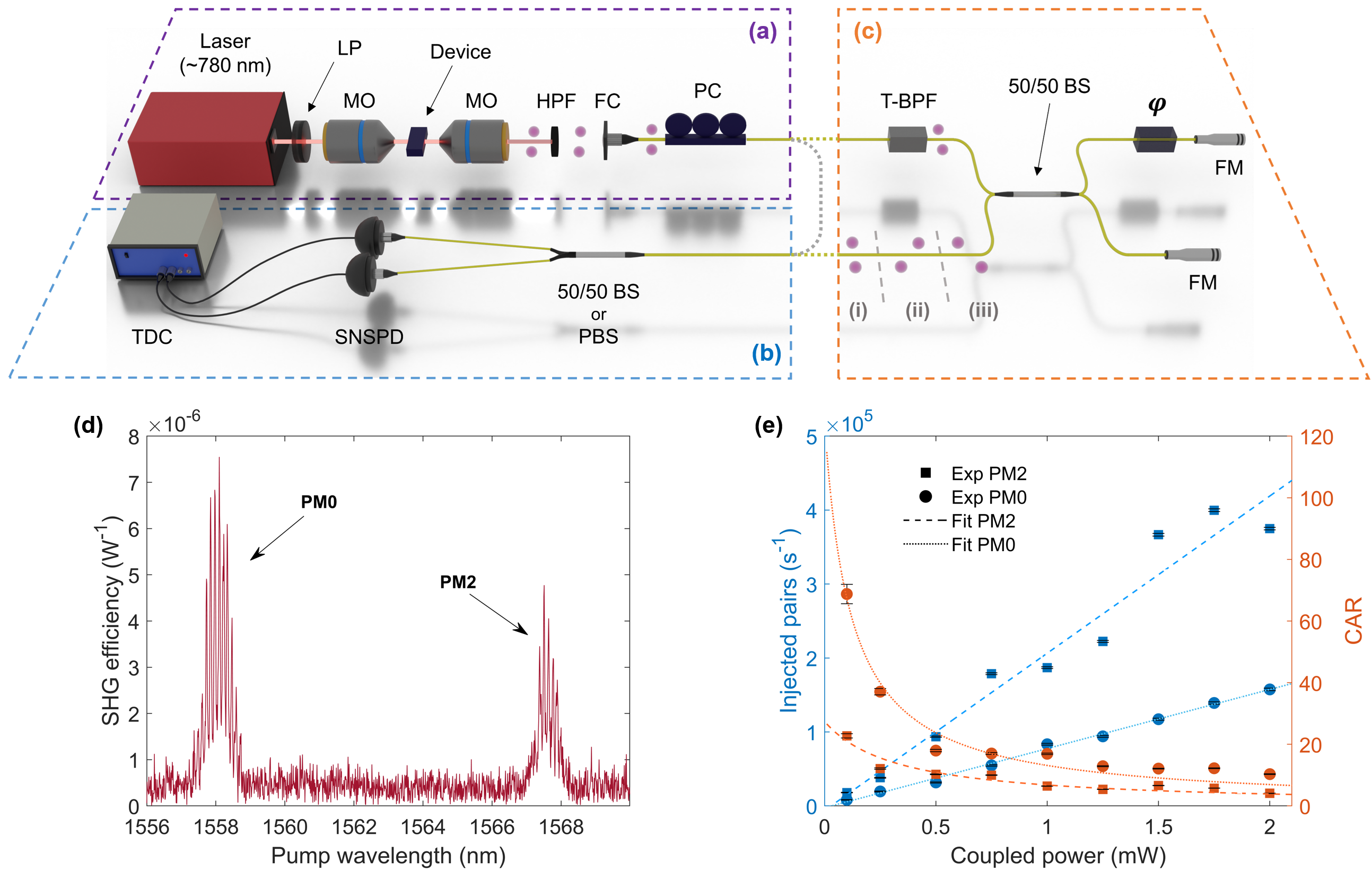}
    \caption{(a,b,c): Sketch of the experimental setup, showing (a) the generation and collection, (b) detection and counting, and (c) Franson interferometer parts of the experiment. Photons produced in (a) are directly sent to (b) for the PGR/CAR measurement (dotted grey connection); (c) is used for the quantum interference measurement (dotted yellow connections). [LP: linear polarizer; MO: microscope objective; HPF: high-pass filter; FC: fiber collimator; PC: polarization controller; T-BPF: tunable band-pass filter; (P)BS: (polarizing) beam splitter; FM: Faraday mirror; SNSPD: superconductive nanowire single-photon detector; TDC: time-to-digital converter.] (d) Measured type 0 and type 2 SHG efficiency in the hybrid device (normalized by the power injected in the silicon waveguide), as a function of the input laser wavelength. The rapid periodic structures are Fabry-Pérot oscillations due to the reflectivity of the facets. (e) Measured internal PGR (in blue) and CAR (in orange) as a function of the coupled optical power: circles for type 0, squares for type 2 phase-matching (PM) process. Curve fittings help visualizing the expected trends. Error bars are calculated assuming a Poissonian statistics of the coincidences counts.}
    \label{fig:setup_SHG_PGR}
\end{figure*}

Second harmonic generation (SHG) measurements are then carried out to investigate the nonlinear response of the hybrid device and precisely determine the experimental resonance frequencies. A CW telecom laser beam is injected into the silicon waveguide through a microscope objective; the SHG signal, at NIR wavelength, is collected from the AlGaAs waveguide through a second microscope objective and measured within a synchronous detection scheme. The input beam is linearly polarized at 45° so as to excite both type 0 and type 2 processes. The result, displayed in Fig. \ref{fig:setup_SHG_PGR}d, shows that both nonlinear processes are retrieved on the same chip with comparable efficiencies, and with resonance wavelengths in the expected spectral range.

\section*{Generation of energy-time entangled photon pairs from the hybrid device}

We now turn to the experiments in the quantum regime, to demonstrate the generation of photon pairs from the hybrid device and assess the quality of the produced quantum state. As sketched in Fig. \ref{fig:setup_SHG_PGR}a, a linearly polarized narrowband CW laser (\textit{TOPTICA}) is coupled to the AlGaAs waveguide by means of a microscope objective. Photon pairs generated by SPDC and transferred into the silicon waveguide are collected from the latter using a second microscope objective, filtered by a long-pass filter (cut-off wavelength 1500 nm, 70 dB rejection) and sent to either a 50/50 beam splitter (BS -- for the type 0 process) or to a polarizing BS (PBS -- for type 2). SPDC photons are finally detected with superconducting nanowire single-photon detectors (SNSPD) connected to a time-to-digital converter (TDC), as shown in Fig. \ref{fig:setup_SHG_PGR}b.

\begin{figure*}[t]
    \centering
    \includegraphics[width=0.85\textwidth]{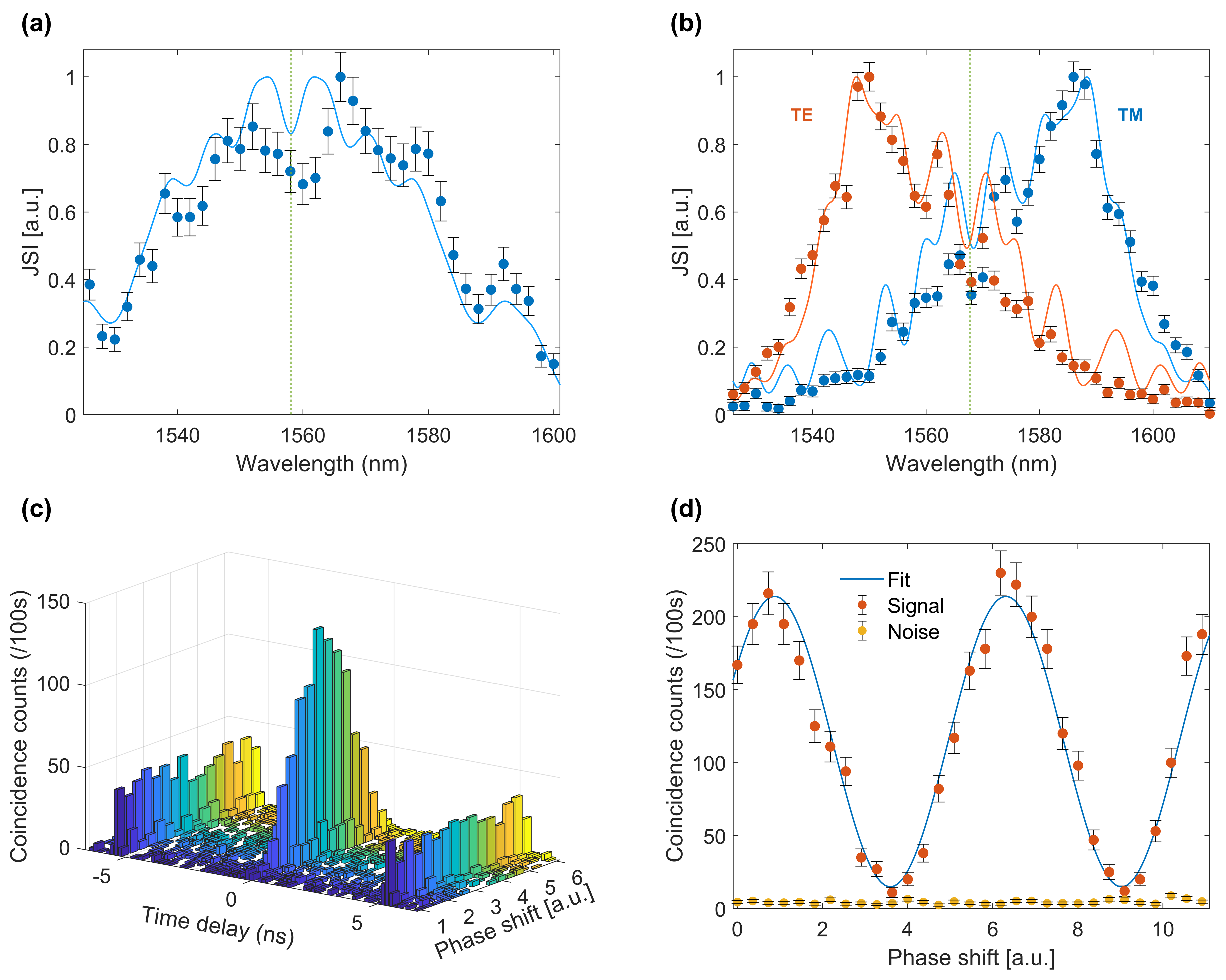}
    \caption{Calculated (plain lines) and measured (points) marginal JSI for (a) type 0 and (b) type 2 transmitted photon pairs; green vertical lines indicate the SPDC degeneracy wavelengths. (c) Recorded coincidence histograms and (d) detail on the central peak as a function of the phase shift for a Franson interferometry measurement. In panels a-b-d the error bars are calculated assuming a Poissonian statistics of the coincidences counts.}
    \label{fig:Franson_JSI}
\end{figure*}

We define the internal pair generation rate (PGR) for the hybrid device as the rate of photon pairs generated in the AlGaAs BRW and transferred to the Si waveguide; this corresponds to the available flux of photon pairs at the end of the coupling region, which is the relevant figure of merit in view of on-chip manipulation in SOI circuits. From the coincidence rates measured at the detectors, we determined this internal PGR and the coincidence-to-accidental ratio (CAR) as a function of the coupled pump power. Considering the measurement setup, the overall system loss – for both arms of the BS – is 15 dB, of which: 4 dB due to the Si waveguide  transmission (losses and facet reflectivity), 6.3 dB to the collection stage, 3.7~dB to the transmission losses of the fibered components and 1 dB to the detection efficiency; additional 3 dB are included when measuring type 0 coincidences, since the two photons are separated in a non-deterministic way by a 50:50 BS. The Klyshko efficiency evaluated as the ratio between true coincidences and single counts is 0.1\% \cite{Atzeni18}. The PGR and CAR as a function of pump power are shown in Fig. \ref{fig:setup_SHG_PGR}e, for the type~0 and type 2 conversion processes: a PGR higher than $10^5$ pairs/s is reached in both cases, with a CAR up to 69 and 23 respectively at low pump power. The fitting curves help in visualizing the expected direct proportionality of the PGR and the inverse proportionality of the CAR with respect to the pump power. 

To demonstrate the capability of the device to reject the optical pump, the same measurements were repeated without any filter at the chip output: the recorded coincidence rate is around 15\% higher (due to a better collection efficiency) while the CAR is only around 30\% lower (Appendix C). Additional measurements (using filters with various cut-off wavelengths) allowed to determine that the decrease of CAR is mainly due to residual luminescence and noise, as most of the accidentals are spectrally distributed above the pump wavelength. The pump beam is thus filtered out with high efficiency in a simple coupling design, without requiring dedicated filters (off- or on-chip), representing an interesting asset compared to alternative recent demonstrations \cite{Sabattoli22,Oser20,kues23}.

We now turn to characterizing the spectral properties of the produced biphoton state. 
The joint spectral intensity (JSI) of the photons generated in the AlGaAs Bragg-reflection waveguide ($\mathcal{C}_{BRW}(\omega_s, \omega_i)$) can be numerically simulated by solving the phase-matching condition for a given conversion process and spectral distribution of the optical pump~\cite{giorgio20}. Assuming a perfectly monochromatic pump, the JSI can be expressed as a function of the signal frequency only. After the mode transfer from the AlGaAs to the SOI waveguide, this marginal JSI $\mathcal{C}(\omega_s)$ can then be written as:
\begin{equation}
    \mathcal{C}^x(\omega_s)=\mathcal{C}^x_{BRW}(\omega_s)\cdot T_u(\omega_s)\cdot T_v(\omega_p-\omega_s),
    \label{eq:marginal_JSI}
\end{equation}
where $x$ is the phase-matching (PM) type (0 or 2), $T(\omega)$ is the device transmission at frequency $\omega$, $u$ and $v$ are the mode polarizations, and we have used the fact that $\omega_i=\omega_p-\omega_s$. For type 0 PM, $u=v=$ TM; for type 2 PM, due to modal birefringence two JSI curves are available, according to the polarization $u$ assigned to the signal photon. The two curves are symmetric with respect to the degeneracy frequency ($\omega_p/2$), as a consequence of the energy conservation. In Fig. \ref{fig:Franson_JSI} we show the calculated $\mathcal{C}$ (solid lines) as a function of the signal wavelength for type 0 (a) and type 2 (b) processes; note that the cavity effects on the transmission spectra were averaged out, for consistency with the experimental measurement. In the same figure, the experimental profiles of $\mathcal{C}$ are also displayed (points). These data were retrieved using the setup in Fig. \ref{fig:setup_SHG_PGR}b: by adding a wavelength-tunable 2 nm band-pass filter on one arm of the (P)BS used to separate the photon pairs, we select the wavelength of the signal photon, and hence that of the idler photon due to their strict correlation in frequency \cite{Kaiser18}.  
For type 2 SPDC (Fig. \ref{fig:Franson_JSI}b) the wavelength offset between TE and TM photons is mainly due to the polarization-dependence of the transmission spectra of the adiabatic coupler. The results show good agreement between the expected and measured marginal JSI curves: this represents a useful tool to analyze the impact of the device conception and fabrication over the produced biphoton state.

Finally, the non-classicality of the biphoton state emitted by the hybrid device is demonstrated via an energy-time entanglement measurement, using a fibered Franson interferometer in the folded configuration \cite{Franson_Original, Folded_Franson}, as sketched in Fig. \ref{fig:setup_SHG_PGR}c. The unbalanced interferometer is composed of a short ($s$) and a long ($l$) arm (path length difference 54 cm), with Faraday mirrors (FM) at their edges to compensate for the polarization rotations occurring within the fibers. The two photons of the pair entering the 50:50 BS can either take different paths (cases (ii) and (iii), $\ket{s_s}\ket{l_i}$ or $\ket{l_s}\ket{s_i}$) or the same one (case (i), $\ket{s_s}\ket{s_i}$ or $\ket{l_s}\ket{l_i}$). 
In the coincidence measurement on the output signal, shown in Fig. \ref{fig:Franson_JSI}c for the type 2 SPDC process, the lateral peaks correspond to the two states belonging to cases (ii) and (iii), while the central peak corresponds to those of case (i). The latter peak thus results from the quantum interference of the post-selected state
\begin{equation}
    \ket\psi=\frac{1}{\sqrt{2}}\left(\ket{s_s}\ket{s_i}+e^{i4\varphi}\ket{l_s}\ket{l_i}\right),
    \label{eq:franson_state}
\end{equation}
where $\varphi$ is the phase shift controlled by a piezoelectric fiber stretcher. 
A 20 nm-large band-pass filter centered at the degeneracy wavelength is placed at the input of the interferometer, in order to narrow the signal bandwidth. This reduces the detrimental contribution of the chromatic dispersion of the interferometer fibers, which introduces a distinguishability between the interfering states that affects the entanglement quality. By varying the phase  $\varphi$, an interference pattern is produced in the central peak, as highlighted in Fig. \ref{fig:Franson_JSI}d, whose visibility quantifies the energy-time entanglement. We obtained a raw (net) visibility of 90.7$\pm$3.0\% (93.8$\pm$3.1\%) for type 0 and  89.1$\pm$1.6\% (92.7$\pm$1.7\%) for type 2 SPDC processes, corresponding to a raw violation of the Bell inequalities by 6.7 and 11.5 standard deviations respectively \cite{Zbinden_Bell}, demonstrating the high entanglement quality of the produced photons and establishing the potentiality of this hybrid device in view of quantum information applications. Even better visibilities (up to 99\%) are obtained upon 5 nm spectral filtering. These results are consistent with the calculated effect of the interferometer chromatic dispersion on the measured visibility.

\section*{Summary and conclusion}

In conclusion, we have demonstrated an hybrid III-V/Silicon quantum photonic device combining the SPDC generation of photon pairs into an AlGaAs waveguide and their vertical routing to a silicon-on-insulator circuitry, thanks to an adiabatic coupling scheme handling both polarization states. This allows the on-chip generation of broadband ($>$ 40 nm, i.e. 50 ITU channels of 100 GHz) telecom photons by type 0 and type 2 SPDC from the hybrid device at room temperature, while the pump beam is strongly rejected without the need for off-chip filtering. Two-photon interference demonstrates the high energy-time entanglement of the produced quantum state, opening the way to diverse applications in quantum information. The demonstrated integration strategy can be generalized to a high number of devices on the same chip, and has the potential to be adapted to CMOS production lines on the model of other recently demonstrated hybrid III-V/Silicon devices \cite{Szelag19}.

In the future, the performances of the demonstrated hybrid device could be further improved by optimizing the design (e.g. using a more complex taper shape) and fabrication process (to gain an even better control over the geometric parameters) to reach a stronger and more polarization-independent coupling of the SPDC photons from the AlGaAs to the Silicon waveguides. We anticipate that a PGR of $0.6\times10^6$ pairs/s for type 0 and $2\times10^6$ pairs/s for type 2 is readily accessible by such improvements. Beyond this, a next promising development is the implementation of the on-chip electrical pumping of the source by monolithically integrating a PIN laser diode into the AlGaAs Bragg-reflection waveguide \cite{Boitier14} part of the device. This would lead to room-temperature, compact and portable photonic chips enabling the electrical injection of photon pairs into high-quality silicon circuits, opening promising avenues for real-world applications in quantum information.

Among the envisioned perspectives, the polarization versatility of our hybrid device is a key asset for protocols based on the polarization degree of freedom.  
Indeed, AlGaAs Bragg-reflection sources enable the generation of polarization-entangled photon pairs with high fidelity \cite{Kang16,Appas21} that, given the very low modal birefringences of both the AlGaAs and SOI parts of the hybrid device ($\Delta n /n <1 $ \%), could be used directly at the chip output, circumventing the usual need of compensating for the group delay between orthogonally polarized photons and opening the way to applications e.g. in multi-user quantum communication networks \cite{Wengerowsky18,Appas21}. In parallel, the large and continuous emission bandwidth of our hybrid source ($>$ 40 nm) could be exploited for frequency entanglement. The latter can be harnessed for diverse applications, to enable high-dimensional quantum information processing tasks \cite{Kues17,Reimer19,Imany19,kues23} or to emulate continuous-variable based protocols, thanks to the direct analogy between the time-frequency variables of a photon pair and the continuous variables of a multiphoton mode of the electromagnetic field \cite{Abouraddy07,Fabre22,Descamps23}. These two features (polarization versatility and large-bandwidth continuous spectrum), together with the simple and direct pump filtering process in our device, constitute a distinctive asset compared to hybrid or monolithic silicon circuits based on microring SFWM sources \cite{Silverstone14,Sabattoli22,Oser20,kues23}.
Both degrees of freedom can also be combined to generate hybrid polarization-frequency entangled photons \cite{Francesconi22} or hyperentangled photon pairs \cite{Kwiat97,Xie15}, opening perspectives e.g. in the field of quantum communication to improve bit rates and resilience to noise \cite{Steinlechner17,Kim21}. Notably, in the continuity of our demonstration these advanced protocols could be realized on-chip thanks to the maturity of the used silicon-on-insulator platform, providing access to a wide array of already demonstrated functionalities such as Mach-Zehnder interferometers, polarizing beam splitters, filters or modulators \cite{Silverstone16}. Combined with the possibility of electrical injection, this opens up the perspective of compact and standalone III-V/Silicon photonic circuits merging the assets of both platforms to progress towards out-of-the-lab implementations of quantum information protocols.

\subsection*{Funding/Acknowledgments}

We thank M. and D. Karr-Ducci for their graphical input, the CIFRE PhD fundings of J. Schuhmann and L. Lazzari with STMicroelectronics, the french RENATECH Network, C. Sciancalepore from SOITEC for providing the SOI wafers, the Plan France 2030 through the projects ANR-22-PETQ-0006 and ANR-22-PETQ-0011, the Paris Ile-de-France Region in the framework of DIM SIRTEQ through the Projects Paris QCI and STARSHIP, the Ville de Paris Emergence LATTICE project, and the European Union’s Horizon Europe research and innovation program under the project "Quantum Security Networks Partnership" (QSNP, grant agreement No 101114043).

%-----------------------------------------------------------------------------------------------------------------------

\section{Appendix A: AlGaAs source epitaxy design}

	The source is designed to optimize the nonlinear conversion efficiency for both type 0 and type 2 SPDC. The AlGaAs structure, grown by molecular beam epitaxy on a GaAs substrate, comprises a 364 nm-thick core made of Al$_{0.45}$Ga$_{0.55}$As positioned in between two Bragg mirrors. The latter consist of alternating 116 nm-thick Al$_{0.25}$Ga$_{0.75}$As layers and 280 nm-thick Al$_{0.8}$Ga$_{0.2}$As layers; the latter parameters are chosen to maximize the nonlinear integral overlap of the parametric process, as follows. The concentration $x$ of the AlGaAs layers is chosen such that the index contrast is maximized (so as to maximize the confinement of the pump mode and thus its overlap with the SPDC modes) while avoiding both the absorption of the pump mode (implying  $x>0.2$) and the oxidation of the layers (implying $x<0.8$). Their thickness is determined by the Bragg reflection condition: $t_i=\lambda / (4 n_i^2-n_{\rm eff}^2)$, where $\lambda$ is the pump wavelength, $n_{\rm eff}$ is the effective index of the pump mode and $n_i$ is the refractive index of layer $i$ having thickness $t_i$. 
	
	\begin{figure*}[!htbp]
		\centering
		\subfigure[]{  \includegraphics[height=4.2cm]{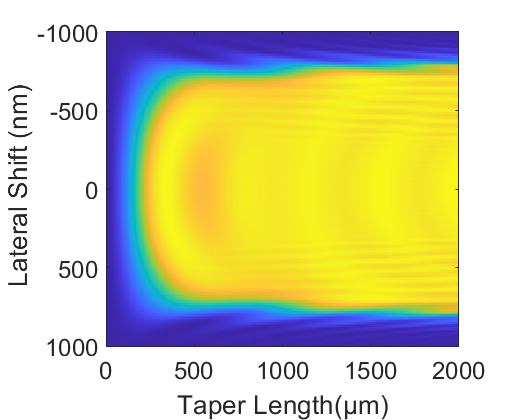}
		}
		\quad    \subfigure[]{
			\includegraphics[height=4.2cm]{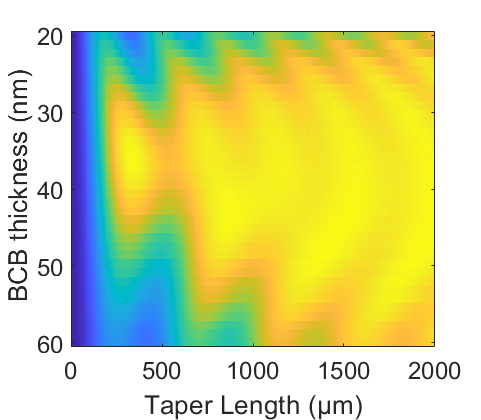}
		}
		\quad    \subfigure[]{
			\includegraphics[height=4.2cm]{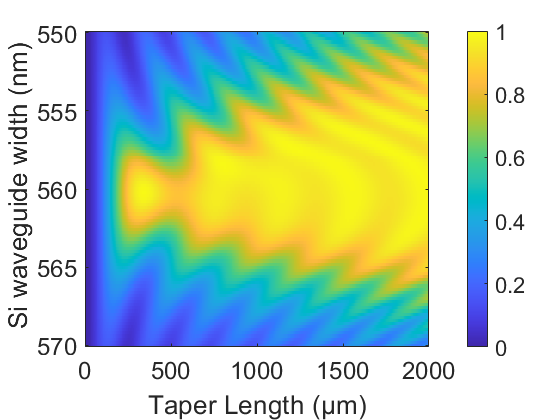}
		}
		\caption{Simulated TM mode transmission as a function of the coupling length and (a) the lateral displacement, (b) the BCB thickness and (c) the silicon waveguide width. Reference conditions are: no lateral shift, 40 nm of BCB thickness, 560 nm of silicon waveguide width.}
		\label{fig:tolerances}
	\end{figure*}

	\begin{figure*}[!htbp]
		\centering
		\includegraphics[width=0.85\textwidth]{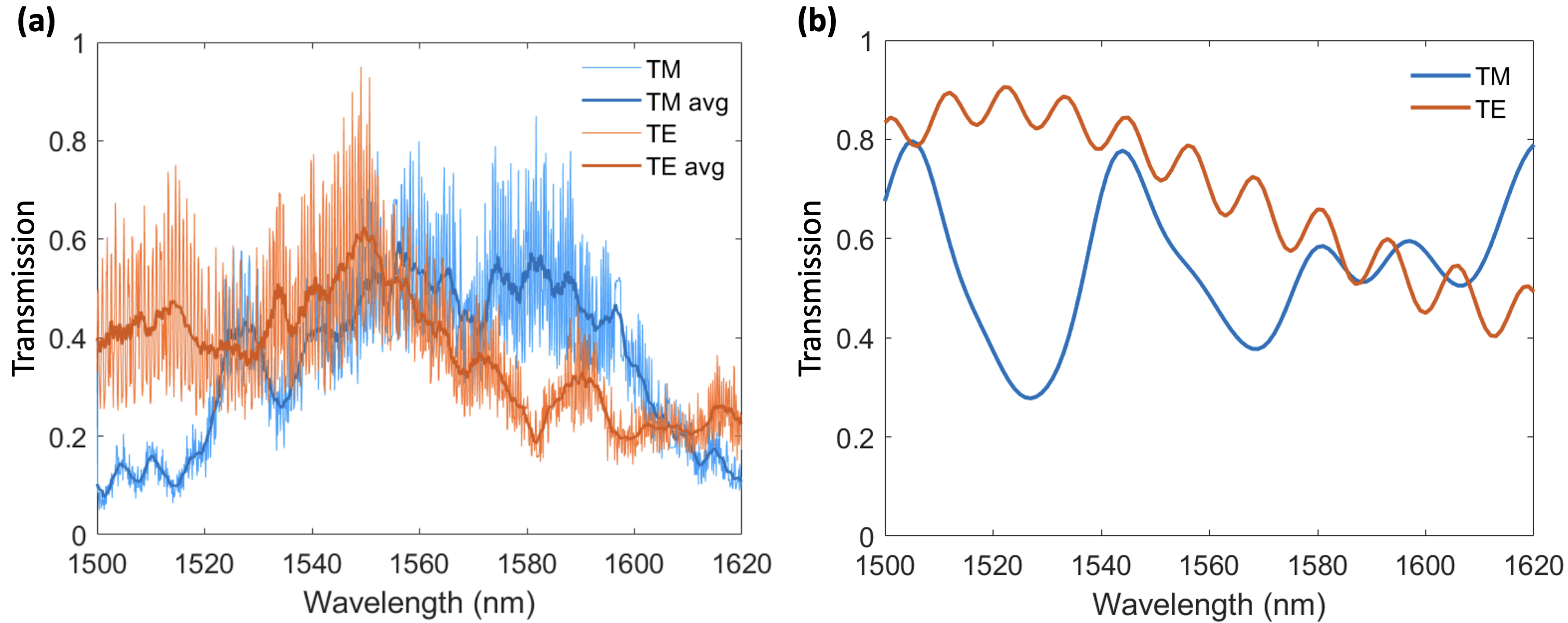}
		\caption{(a) Measured TM and TE transmission spectra of the hybrid device; cavity oscillations are averaged out in the superimposed curves (bold lines). (b) Numerically simulated transmission spectra.}
		\label{Transmission}
	\end{figure*}

\section{Appendix B: Vertical adiabatic coupling}

\setcounter{equation}{0}
\renewcommand{\theequation}{B\arabic{equation}}

The hybrid structure is designed to achieve the optical mode transfer from the AlGaAs waveguide to the silicon one via adiabatic evanescent coupling. The device is fabricated in such a way that the two waveguides are superimposed and aligned along the longitudinal direction, corresponding to the mode propagation. The two waveguides are close enough so that the evanescent tails of the guided modes overlap: exploiting coupled mode theory \cite{Coupled_Mode_Theory}, we can predict and control the leaking of a mode into the other waveguide mode, in order to obtain efficient optical power transmission while minimizing the footprint and ensuring an acceptable robustness and manufacturability. 

Considering two waveguides with propagation constants $\beta_1$ and $\beta_2$, a coherent and even superposition of their modes is an even supermode that can be written as:
\begin{equation}
	\Tilde{E}_e(x,y,z)=\frac{1}{\sqrt{2}}\begin{bmatrix} \sqrt{1-\delta/S}\\\sqrt{1+\delta/S} \end{bmatrix}e^{-i\beta_ex},
	\label{eq:supermode}
\end{equation}
with $x$ the propagation direction, $\delta=(\beta_2-\beta_1)/2$, $S=\sqrt{\delta^2+\kappa^2}$, where $\kappa$ is the coupling strength, and $\beta_e=(\beta_2+\beta_1)/2+S$ is the propagation constant of the supermode. The coupling strength $\kappa$ is proportional to the overlap integral of the two transverse mode distributions; for given confined modes, it significantly decreases with the distance between the two waveguides, which is thus a crucial parameter. For this reason, the AlGaAs waveguides feature only two Bragg mirrors on the side closer to the SOI platform. The two components of the vector \ref{eq:supermode} represent the weight of each transverse mode distribution in the supermode and, hence, the amount of optical power confined in each waveguide. By varying the propagation constants $\beta_1$ and $\beta_2$, one can control the localization of the optical power. The propagation constant being proportional to the effective refractive index, the easiest way to modify it is to act on the waveguide dimensions: in particular, the width can be changed effortlessly at the fabrication stage. 

For this work, we designed and fabricated a linear taper in the AlGaAs waveguide. This configuration ensures the level of robustness required for facing the finite fabrication precision and eventual processing imperfections. Numerical simulations performed with a commercial Finite Difference Eigenmode (FDE) solver show how the presented device is, in particular, robust against the relative lateral shift between the superimposed waveguides (up to 500 nm shift, Fig. \ref{fig:tolerances}a), the BCB thickness variation (up to 10 nm, Fig. \ref{fig:tolerances}b) and the silicon waveguide width (up to 5 nm, Fig. \ref{fig:tolerances}c). The device tolerance increases with the taper length, as expected for an adiabatic coupling design.

\section{Appendix C: Device transmission and pump rejection}

For characterizing the device transmission in the telecom range, a linearly polarized CW laser (\textit{Tunics}) is coupled into the AlGaAs waveguide by means of a microscope objective. The radiation at the output of the silicon waveguide is collected with another objective. Free-space linear polarizers are used to set the input polarization and check the output one. Input and output optical powers are measured with a powermeter; from these values, considering the estimated injection and collection efficiency, the facets' reflectivities (numerically simulated with a commercial FDTD solver) and the waveguides' optical losses, the actual mode coupling efficiency can be retrieved; straight AlGaAs and silicon waveguides (alone, without tapers) are fabricated on the same chip, in order to measure the respective optical losses with the same setup, using the Fabry-Pérot fringes method \cite{FP}. The experimental coupling efficiency from the AlGaAs to the SOI waveguide is displayed in Fig. \ref{Transmission}a for both polarizations. The oscillations in the transmitted power are due to cavity effects caused by the modal reflection at the facets; bold lines show the transmission curves after averaging out these rapid oscillations.
For comparison, Fig. \ref{Transmission}b shows the numerically simulated transmission spectra for TE and TM polarizations. The simulated transmission spectrum for the TE is peaked towards lower wavelengths, as for the measured data; for the TM mode, the simulated transmission is peaked around 1550 nm, similarly to the experiment, but it also shows oscillations whose amplitude is higher than in the experiment. This discrepancy is likely due to a slight difference between the nominal structure considered for the simulation and the effectively realized structure with its inevitable fabrication imperfections.

\begin{figure}[!htbp]
	\centering
	\includegraphics[width=0.42\textwidth]{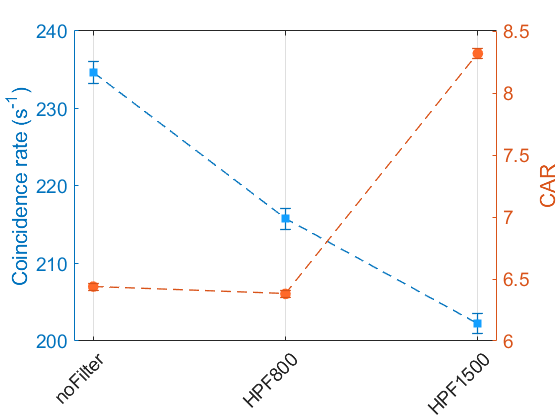}
	\caption{Type 2 coincidence rates and CAR of the hybrid device at a coupled power of 1.5 mW for different spectral filtering configurations of the output signal: no filtering, high-pass filtering with cut-off wavelength at 800 nm (HPF800) and at 1500 nm (HPF1500); points are connected to help the data visualization.}
	\label{Noise}
\end{figure}

A distinctive asset of this hybrid structure lies in the intrinsic suppression of the optical pump. This is obtained thanks to two reasons: first, the adiabatic coupling is inefficient outside the designed spectral range; second, silicon would absorb any residual transmission of the pump beam (775 nm) into the silicon waveguide.  To demonstrate the pump rejection capability of the device, we repeated the coincidence rate measurements without the high-pass filter on the output signal (cut-off wavelength 1500 nm). In this condition, when collecting photon pairs from a monolithic AlGaAs waveguide, we would saturate the detectors; to perform a coincidence measurement, we typically have to introduce 70 dB of additional rejection (20 dB with a dichroic mirror, 50 dB with the high-pass filter) to suppress the pump beam. By contrast, with the hybrid device the coincidence signal can be retrieved even without such off-chip filtering: the recorded coincidence rate, shown in Fig. \ref{Noise}, is around 15\% higher, while the CAR is around 30\% reduced. The increase in coincidence counts is ascribable to the absence of the filter transmission contribution, while the increase in accidentals is mainly due to the waveguide luminescence: most of the accidentals are spectrally distributed above the pump wavelength, as demonstrated using a high-pass filter with cut-off wavelength of 800 nm.

\section{Appendix D: Franson interferometric setup and measurements}

\setcounter{equation}{0}
\renewcommand{\theequation}{D\arabic{equation}}

To observe quantum interference in a Franson measurement, the interferometer must be carefully designed. A crucial parameter to obtain the superposition state of Eq. (3) of the main text, is the traveling time difference between the two arms, $\Delta T=\Delta Ln_{\rm eff}/c$, with $\Delta L$ the path unbalance and $n_{\rm eff}$ the effective refractive index of the fiber. $\Delta T$ must be much larger than the coherence time of each photon, to avoid single-photon interference, and than the combined jitter of the detector and TDC, to experimentally distinguish the coincidence peaks in case (i) from the one in case (ii) [see Fig. 2b of the main text]; moreover, in order to have two-photon interference, $\Delta T$ must be much shorter than the pump coherence time \cite{Franson_Original}. In our case, we set $\Delta L=54$ cm, satisfying all the requirements. Another important parameter for a fibered interferometer is the temperature stabilization: thermal phase drifts induced by instabilities strongly affect the interference visibility for a given integration time \cite{KAISER20147}. By thermally isolating the setup, we managed to extend this integration time up to 10 minutes, with negligible deterioration of the visibility.
The used fibers are not polarization-maintaining, since any polarization rotation occurring while propagating back-and-forth in the interferometer is compensated by the Faraday mirrors. Only in the case in which the introduced shift varies within a single round trip would such compensation not be effective: this case is, however, highly unlikely, considering the time scales of mechanical and thermal drifts (roughly going from 0.01 Hz to 10 kHz) and the round trip time (around  5 ns). The photons are then separated via a (P)BS immediately after exiting the interferometer, so that no further degradation due to polarization drift is introduced in the quantum state.

The visibility is obtained by fitting the experimental data to the function
\begin{equation}
	f(x)=A\left[1+V\cos\left((x-x_0)\frac{\pi}{p}\right)\right],
	\label{eq:fitting}
\end{equation}
where $A$ is the oscillation amplitude, $V$ the curve visibility, $x_0$ the phase offset and $p$ the oscillation periodicity. The Poissonian noise characterizing the coincidence measurement (error bars in Fig. 3d of the main text) is taken into account by assigning to each data point a weight that is inversely proportional to its associated statistical variance. Raw visibilities of (90.7$\pm$3.0)\% and (89.1$\pm$1.6)\% are obtained for type 0 and type 2 SPDC photons respectively; after subtracting the noise contribution given by accidental counts, we obtain the net visibilities reported in the article. Under the assumption that the photon pairs going through the other arm of the BS (not measured here) show the same interference visibility, we can retrieve the Bell parameter $S$ \cite{Zbinden_Bell}:
\begin{equation}
	S=2\sqrt{2}V.
	\label{eq:fitting}
\end{equation}
Therefore, while for classical fields the correlation visibility cannot exceed 50\% \cite{Franson_vis}, visibilities larger than $1/\sqrt{2}\simeq70.7$\% correspond to a violation of the Bell inequalities, witnessing the non-local nature of the measured entanglement.

As a last remark, for a fibered -- or, in general, non free-space -- interferometer the visibility is limited by the chromatic dispersion of the fibers, which introduces a degree of distinguishability between the interfering states. This effect is related to the interferometer unbalance $\Delta L$ and to the signal bandwidth. By using a spectral filter, we can control (and reduce) the bandwidth of the generated quantum state. In particular, for a spectral width of 20 nm, as in the case of the displayed measurements, we estimate the visibility to be limited to around 92\%, which is consistent with our experimental results. Higher visibilities can be targeted with narrower filtering or shorter $\Delta L$, by using non-dispersive fibers in the Franson interferometer.

%\bibliography{Biblio.bib}

\begin{thebibliography}{65}%
	\makeatletter
	\providecommand \@ifxundefined [1]{%
		\@ifx{#1\undefined}
	}%
	\providecommand \@ifnum [1]{%
		\ifnum #1\expandafter \@firstoftwo
		\else \expandafter \@secondoftwo
		\fi
	}%
	\providecommand \@ifx [1]{%
		\ifx #1\expandafter \@firstoftwo
		\else \expandafter \@secondoftwo
		\fi
	}%
	\providecommand \natexlab [1]{#1}%
	\providecommand \enquote  [1]{``#1''}%
	\providecommand \bibnamefont  [1]{#1}%
	\providecommand \bibfnamefont [1]{#1}%
	\providecommand \citenamefont [1]{#1}%
	\providecommand \href@noop [0]{\@secondoftwo}%
	\providecommand \href [0]{\begingroup \@sanitize@url \@href}%
	\providecommand \@href[1]{\@@startlink{#1}\@@href}%
	\providecommand \@@href[1]{\endgroup#1\@@endlink}%
	\providecommand \@sanitize@url [0]{\catcode `\\12\catcode `\$12\catcode
		`\&12\catcode `\#12\catcode `\^12\catcode `\_12\catcode `\%12\relax}%
	\providecommand \@@startlink[1]{}%
	\providecommand \@@endlink[0]{}%
	\providecommand \url  [0]{\begingroup\@sanitize@url \@url }%
	\providecommand \@url [1]{\endgroup\@href {#1}{\urlprefix }}%
	\providecommand \urlprefix  [0]{URL }%
	\providecommand \Eprint [0]{\href }%
	\providecommand \doibase [0]{https://doi.org/}%
	\providecommand \selectlanguage [0]{\@gobble}%
	\providecommand \bibinfo  [0]{\@secondoftwo}%
	\providecommand \bibfield  [0]{\@secondoftwo}%
	\providecommand \translation [1]{[#1]}%
	\providecommand \BibitemOpen [0]{}%
	\providecommand \bibitemStop [0]{}%
	\providecommand \bibitemNoStop [0]{.\EOS\space}%
	\providecommand \EOS [0]{\spacefactor3000\relax}%
	\providecommand \BibitemShut  [1]{\csname bibitem#1\endcsname}%
	\let\auto@bib@innerbib\@empty
	%</preamble>
	\bibitem [{\citenamefont {Flamini}\ \emph {et~al.}(2018)\citenamefont
		{Flamini}, \citenamefont {Spagnolo},\ and\ \citenamefont
		{Sciarrino}}]{Flamini19}%
	\BibitemOpen
	\bibfield  {author} {\bibinfo {author} {\bibfnamefont {F.}~\bibnamefont
			{Flamini}}, \bibinfo {author} {\bibfnamefont {N.}~\bibnamefont {Spagnolo}},\
		and\ \bibinfo {author} {\bibfnamefont {F.}~\bibnamefont {Sciarrino}},\
	}\bibfield  {title} {\bibinfo {title} {Photonic quantum information
			processing: a review},\ }\href@noop {} {\bibfield  {journal} {\bibinfo
			{journal} {Reports on Progress in Physics}\ }\textbf {\bibinfo {volume}
			{82}},\ \bibinfo {pages} {016001} (\bibinfo {year} {2018})}\BibitemShut
	{NoStop}%
	\bibitem [{\citenamefont {Wang}\ \emph {et~al.}(2020)\citenamefont {Wang},
		\citenamefont {Sciarrino}, \citenamefont {Laing},\ and\ \citenamefont
		{Thompson}}]{Wang20}%
	\BibitemOpen
	\bibfield  {author} {\bibinfo {author} {\bibfnamefont {J.}~\bibnamefont
			{Wang}}, \bibinfo {author} {\bibfnamefont {F.}~\bibnamefont {Sciarrino}},
		\bibinfo {author} {\bibfnamefont {A.}~\bibnamefont {Laing}},\ and\ \bibinfo
		{author} {\bibfnamefont {M.~G.}\ \bibnamefont {Thompson}},\ }\bibfield
	{title} {\bibinfo {title} {Integrated photonic quantum technologies},\
	}\href@noop {} {\bibfield  {journal} {\bibinfo  {journal} {Nature Photonics}\
		}\textbf {\bibinfo {volume} {14}},\ \bibinfo {pages} {273} (\bibinfo {year}
		{2020})}\BibitemShut {NoStop}%
	\bibitem [{\citenamefont {Pelucchi}\ \emph {et~al.}(2022)\citenamefont
		{Pelucchi}, \citenamefont {Fagas}, \citenamefont {Aharonovich}, \citenamefont
		{Englund}, \citenamefont {Figueroa}, \citenamefont {Gong}, \citenamefont
		{Hannes}, \citenamefont {Liu}, \citenamefont {Lu}, \citenamefont {Matsuda},
		\citenamefont {Pan}, \citenamefont {Schreck}, \citenamefont {Sciarrino},
		\citenamefont {Silberhorn}, \citenamefont {Wang},\ and\ \citenamefont
		{J{\"o}ns}}]{Pelucchi22}%
	\BibitemOpen
	\bibfield  {author} {\bibinfo {author} {\bibfnamefont {E.}~\bibnamefont
			{Pelucchi}}, \bibinfo {author} {\bibfnamefont {G.}~\bibnamefont {Fagas}},
		\bibinfo {author} {\bibfnamefont {I.}~\bibnamefont {Aharonovich}}, \bibinfo
		{author} {\bibfnamefont {D.}~\bibnamefont {Englund}}, \bibinfo {author}
		{\bibfnamefont {E.}~\bibnamefont {Figueroa}}, \bibinfo {author}
		{\bibfnamefont {Q.}~\bibnamefont {Gong}}, \bibinfo {author} {\bibfnamefont
			{H.}~\bibnamefont {Hannes}}, \bibinfo {author} {\bibfnamefont
			{J.}~\bibnamefont {Liu}}, \bibinfo {author} {\bibfnamefont {C.-Y.}\
			\bibnamefont {Lu}}, \bibinfo {author} {\bibfnamefont {N.}~\bibnamefont
			{Matsuda}}, \bibinfo {author} {\bibfnamefont {J.-W.}\ \bibnamefont {Pan}},
		\bibinfo {author} {\bibfnamefont {F.}~\bibnamefont {Schreck}}, \bibinfo
		{author} {\bibfnamefont {F.}~\bibnamefont {Sciarrino}}, \bibinfo {author}
		{\bibfnamefont {C.}~\bibnamefont {Silberhorn}}, \bibinfo {author}
		{\bibfnamefont {J.}~\bibnamefont {Wang}},\ and\ \bibinfo {author}
		{\bibfnamefont {K.~D.}\ \bibnamefont {J{\"o}ns}},\ }\bibfield  {title}
	{\bibinfo {title} {The potential and global outlook of integrated photonics
			for quantum technologies},\ }\href@noop {} {\bibfield  {journal} {\bibinfo
			{journal} {Nature Reviews Physics}\ }\textbf {\bibinfo {volume} {4}},\
		\bibinfo {pages} {194} (\bibinfo {year} {2022})}\BibitemShut {NoStop}%
	\bibitem [{\citenamefont {Silverstone}\ \emph {et~al.}(2016)\citenamefont
		{Silverstone}, \citenamefont {Wang}, \citenamefont {Bonneau}, \citenamefont
		{Sibson}, \citenamefont {Santagati}, \citenamefont {Erven}, \citenamefont
		{O'Brien},\ and\ \citenamefont {Thompson}}]{Silverstone16}%
	\BibitemOpen
	\bibfield  {author} {\bibinfo {author} {\bibfnamefont {J.~W.}\ \bibnamefont
			{Silverstone}}, \bibinfo {author} {\bibfnamefont {J.}~\bibnamefont {Wang}},
		\bibinfo {author} {\bibfnamefont {D.}~\bibnamefont {Bonneau}}, \bibinfo
		{author} {\bibfnamefont {P.}~\bibnamefont {Sibson}}, \bibinfo {author}
		{\bibfnamefont {R.}~\bibnamefont {Santagati}}, \bibinfo {author}
		{\bibfnamefont {C.}~\bibnamefont {Erven}}, \bibinfo {author} {\bibfnamefont
			{J.}~\bibnamefont {O'Brien}},\ and\ \bibinfo {author} {\bibfnamefont
			{M.}~\bibnamefont {Thompson}},\ }\bibfield  {title} {\bibinfo {title}
		{Silicon quantum photonics},\ }\href@noop {} {\bibfield  {journal} {\bibinfo
			{journal} {2016 International Conference on Optical MEMS and Nanophotonics}\
		}\textbf {\bibinfo {volume} {1}},\ \bibinfo {pages} {1} (\bibinfo {year}
		{2016})}\BibitemShut {NoStop}%
	\bibitem [{\citenamefont {Dietrich}\ \emph {et~al.}(2016)\citenamefont
		{Dietrich}, \citenamefont {Fiore}, \citenamefont {Thompson}, \citenamefont
		{Kamp},\ and\ \citenamefont {H{\"o}fling}}]{Dietrich16}%
	\BibitemOpen
	\bibfield  {author} {\bibinfo {author} {\bibfnamefont {C.~P.}\ \bibnamefont
			{Dietrich}}, \bibinfo {author} {\bibfnamefont {A.}~\bibnamefont {Fiore}},
		\bibinfo {author} {\bibfnamefont {M.~G.}\ \bibnamefont {Thompson}}, \bibinfo
		{author} {\bibfnamefont {M.}~\bibnamefont {Kamp}},\ and\ \bibinfo {author}
		{\bibfnamefont {S.}~\bibnamefont {H{\"o}fling}},\ }\bibfield  {title}
	{\bibinfo {title} {{GaAs} integrated quantum photonics: Towards compact and
			multi-functional quantum photonic integrated circuits},\ }\href@noop {}
	{\bibfield  {journal} {\bibinfo  {journal} {Laser \& Photonics Reviews}\
		}\textbf {\bibinfo {volume} {10}},\ \bibinfo {pages} {870} (\bibinfo {year}
		{2016})}\BibitemShut {NoStop}%
	\bibitem [{\citenamefont {Alibart}\ \emph {et~al.}(2016)\citenamefont
		{Alibart}, \citenamefont {D’Auria}, \citenamefont {De~Micheli},
		\citenamefont {Doutre}, \citenamefont {Kaiser}, \citenamefont {Labont{\'e}},
		\citenamefont {Lunghi}, \citenamefont {Picholle},\ and\ \citenamefont
		{Tanzilli}}]{Alibart16}%
	\BibitemOpen
	\bibfield  {author} {\bibinfo {author} {\bibfnamefont {O.}~\bibnamefont
			{Alibart}}, \bibinfo {author} {\bibfnamefont {V.}~\bibnamefont {D’Auria}},
		\bibinfo {author} {\bibfnamefont {M.}~\bibnamefont {De~Micheli}}, \bibinfo
		{author} {\bibfnamefont {F.}~\bibnamefont {Doutre}}, \bibinfo {author}
		{\bibfnamefont {F.}~\bibnamefont {Kaiser}}, \bibinfo {author} {\bibfnamefont
			{L.}~\bibnamefont {Labont{\'e}}}, \bibinfo {author} {\bibfnamefont
			{T.}~\bibnamefont {Lunghi}}, \bibinfo {author} {\bibfnamefont
			{{\'E}.}~\bibnamefont {Picholle}},\ and\ \bibinfo {author} {\bibfnamefont
			{S.}~\bibnamefont {Tanzilli}},\ }\bibfield  {title} {\bibinfo {title}
		{Quantum photonics at telecom wavelengths based on lithium niobate
			waveguides},\ }\href@noop {} {\bibfield  {journal} {\bibinfo  {journal}
			{Journal of Optics}\ }\textbf {\bibinfo {volume} {18}},\ \bibinfo {pages}
		{104001} (\bibinfo {year} {2016})}\BibitemShut {NoStop}%
	\bibitem [{\citenamefont {Elshaari}\ \emph {et~al.}(2020)\citenamefont
		{Elshaari}, \citenamefont {Pernice}, \citenamefont {Srinivasan},
		\citenamefont {Benson},\ and\ \citenamefont {Zwiller}}]{Elshaari20}%
	\BibitemOpen
	\bibfield  {author} {\bibinfo {author} {\bibfnamefont {A.~W.}\ \bibnamefont
			{Elshaari}}, \bibinfo {author} {\bibfnamefont {W.}~\bibnamefont {Pernice}},
		\bibinfo {author} {\bibfnamefont {K.}~\bibnamefont {Srinivasan}}, \bibinfo
		{author} {\bibfnamefont {O.}~\bibnamefont {Benson}},\ and\ \bibinfo {author}
		{\bibfnamefont {V.}~\bibnamefont {Zwiller}},\ }\bibfield  {title} {\bibinfo
		{title} {Hybrid integrated quantum photonic circuits},\ }\href@noop {}
	{\bibfield  {journal} {\bibinfo  {journal} {Nature Photonics}\ }\textbf
		{\bibinfo {volume} {14}},\ \bibinfo {pages} {285} (\bibinfo {year}
		{2020})}\BibitemShut {NoStop}%
	\bibitem [{\citenamefont {Kim}\ \emph {et~al.}(2020)\citenamefont {Kim},
		\citenamefont {Aghaeimeibodi}, \citenamefont {Carolan}, \citenamefont
		{Englund},\ and\ \citenamefont {Waks}}]{Kim20}%
	\BibitemOpen
	\bibfield  {author} {\bibinfo {author} {\bibfnamefont {J.-H.}\ \bibnamefont
			{Kim}}, \bibinfo {author} {\bibfnamefont {S.}~\bibnamefont {Aghaeimeibodi}},
		\bibinfo {author} {\bibfnamefont {J.}~\bibnamefont {Carolan}}, \bibinfo
		{author} {\bibfnamefont {D.}~\bibnamefont {Englund}},\ and\ \bibinfo {author}
		{\bibfnamefont {E.}~\bibnamefont {Waks}},\ }\bibfield  {title} {\bibinfo
		{title} {Hybrid integration methods for on-chip quantum photonics},\
	}\href@noop {} {\bibfield  {journal} {\bibinfo  {journal} {Optica}\ }\textbf
		{\bibinfo {volume} {7}},\ \bibinfo {pages} {291} (\bibinfo {year}
		{2020})}\BibitemShut {NoStop}%
	\bibitem [{\citenamefont {Siew}\ \emph {et~al.}(2021)\citenamefont {Siew},
		\citenamefont {Li}, \citenamefont {Gao}, \citenamefont {Zheng}, \citenamefont
		{Zhang}, \citenamefont {Guo}, \citenamefont {Xie}, \citenamefont {Song},
		\citenamefont {Dong}, \citenamefont {Luo}, \citenamefont {Li}, \citenamefont
		{Luo},\ and\ \citenamefont {Lo}}]{Review_Siew}%
	\BibitemOpen
	\bibfield  {author} {\bibinfo {author} {\bibfnamefont {S.~Y.}\ \bibnamefont
			{Siew}}, \bibinfo {author} {\bibfnamefont {B.}~\bibnamefont {Li}}, \bibinfo
		{author} {\bibfnamefont {F.}~\bibnamefont {Gao}}, \bibinfo {author}
		{\bibfnamefont {H.~Y.}\ \bibnamefont {Zheng}}, \bibinfo {author}
		{\bibfnamefont {W.}~\bibnamefont {Zhang}}, \bibinfo {author} {\bibfnamefont
			{P.}~\bibnamefont {Guo}}, \bibinfo {author} {\bibfnamefont {S.~W.}\
			\bibnamefont {Xie}}, \bibinfo {author} {\bibfnamefont {A.}~\bibnamefont
			{Song}}, \bibinfo {author} {\bibfnamefont {B.}~\bibnamefont {Dong}}, \bibinfo
		{author} {\bibfnamefont {L.~W.}\ \bibnamefont {Luo}}, \bibinfo {author}
		{\bibfnamefont {C.}~\bibnamefont {Li}}, \bibinfo {author} {\bibfnamefont
			{X.}~\bibnamefont {Luo}},\ and\ \bibinfo {author} {\bibfnamefont {G.-Q.}\
			\bibnamefont {Lo}},\ }\bibfield  {title} {\bibinfo {title} {Review of silicon
			photonics technology and platform development},\ }\href@noop {} {\bibfield
		{journal} {\bibinfo  {journal} {Journal of Lightwave Technology}\ }\textbf
		{\bibinfo {volume} {39}},\ \bibinfo {pages} {4374} (\bibinfo {year}
		{2021})}\BibitemShut {NoStop}%
	\bibitem [{\citenamefont {Baboux}\ \emph {et~al.}(2023)\citenamefont {Baboux},
		\citenamefont {Moody},\ and\ \citenamefont {Ducci}}]{Baboux23}%
	\BibitemOpen
	\bibfield  {author} {\bibinfo {author} {\bibfnamefont {F.}~\bibnamefont
			{Baboux}}, \bibinfo {author} {\bibfnamefont {G.}~\bibnamefont {Moody}},\ and\
		\bibinfo {author} {\bibfnamefont {S.}~\bibnamefont {Ducci}},\ }\bibfield
	{title} {\bibinfo {title} {Nonlinear integrated quantum photonics with
			{AlGaAs}},\ }\href@noop {} {\bibfield  {journal} {\bibinfo  {journal}
			{Optica}\ }\textbf {\bibinfo {volume} {10}},\ \bibinfo {pages} {917}
		(\bibinfo {year} {2023})}\BibitemShut {NoStop}%
	\bibitem [{\citenamefont {Liang}\ and\ \citenamefont {Bowers}(2010)}]{Liang10}%
	\BibitemOpen
	\bibfield  {author} {\bibinfo {author} {\bibfnamefont {D.}~\bibnamefont
			{Liang}}\ and\ \bibinfo {author} {\bibfnamefont {J.~E.}\ \bibnamefont
			{Bowers}},\ }\bibfield  {title} {\bibinfo {title} {Recent progress in lasers
			on silicon},\ }\href@noop {} {\bibfield  {journal} {\bibinfo  {journal}
			{Nature Photonics}\ }\textbf {\bibinfo {volume} {4}},\ \bibinfo {pages} {511}
		(\bibinfo {year} {2010})}\BibitemShut {NoStop}%
	\bibitem [{\citenamefont {Davanco}\ \emph {et~al.}(2017)\citenamefont
		{Davanco}, \citenamefont {Liu}, \citenamefont {Sapienza}, \citenamefont
		{Zhang}, \citenamefont {Cardoso}, \citenamefont {Verma}, \citenamefont
		{Mirin}, \citenamefont {Nam}, \citenamefont {Liu},\ and\ \citenamefont
		{Srinivasan}}]{Davanco17}%
	\BibitemOpen
	\bibfield  {author} {\bibinfo {author} {\bibfnamefont {M.}~\bibnamefont
			{Davanco}}, \bibinfo {author} {\bibfnamefont {J.}~\bibnamefont {Liu}},
		\bibinfo {author} {\bibfnamefont {L.}~\bibnamefont {Sapienza}}, \bibinfo
		{author} {\bibfnamefont {C.-Z.}\ \bibnamefont {Zhang}}, \bibinfo {author}
		{\bibfnamefont {J.~V. D.~M.}\ \bibnamefont {Cardoso}}, \bibinfo {author}
		{\bibfnamefont {V.}~\bibnamefont {Verma}}, \bibinfo {author} {\bibfnamefont
			{R.}~\bibnamefont {Mirin}}, \bibinfo {author} {\bibfnamefont {S.~W.}\
			\bibnamefont {Nam}}, \bibinfo {author} {\bibfnamefont {L.}~\bibnamefont
			{Liu}},\ and\ \bibinfo {author} {\bibfnamefont {K.}~\bibnamefont
			{Srinivasan}},\ }\bibfield  {title} {\bibinfo {title} {Heterogeneous
			integration for on-chip quantum photonic circuits with single quantum dot
			devices},\ }\href@noop {} {\bibfield  {journal} {\bibinfo  {journal} {Nature
				Communications}\ }\textbf {\bibinfo {volume} {8}},\ \bibinfo {pages} {1}
		(\bibinfo {year} {2017})}\BibitemShut {NoStop}%
	\bibitem [{\citenamefont {Yue}\ \emph {et~al.}(2018)\citenamefont {Yue},
		\citenamefont {Dou}, \citenamefont {Wang}, \citenamefont {Ma}, \citenamefont
		{Niu},\ and\ \citenamefont {Sun}}]{Yue18}%
	\BibitemOpen
	\bibfield  {author} {\bibinfo {author} {\bibfnamefont {P.}~\bibnamefont
			{Yue}}, \bibinfo {author} {\bibfnamefont {X.}~\bibnamefont {Dou}}, \bibinfo
		{author} {\bibfnamefont {H.}~\bibnamefont {Wang}}, \bibinfo {author}
		{\bibfnamefont {B.}~\bibnamefont {Ma}}, \bibinfo {author} {\bibfnamefont
			{Z.}~\bibnamefont {Niu}},\ and\ \bibinfo {author} {\bibfnamefont
			{B.}~\bibnamefont {Sun}},\ }\bibfield  {title} {\bibinfo {title} {{Single
				photon emissions from InAs/GaAs quantum dots embedded in GaAs/SiO$_2$ hybrid
				microdisks}},\ }\href@noop {} {\bibfield  {journal} {\bibinfo  {journal}
			{Optics Communications}\ }\textbf {\bibinfo {volume} {411}},\ \bibinfo
		{pages} {114} (\bibinfo {year} {2018})}\BibitemShut {NoStop}%
	\bibitem [{\citenamefont {Osada}\ \emph {et~al.}(2019)\citenamefont {Osada},
		\citenamefont {Ota}, \citenamefont {Katsumi}, \citenamefont {Kakuda},
		\citenamefont {Iwamoto},\ and\ \citenamefont {Arakawa}}]{Osada19}%
	\BibitemOpen
	\bibfield  {author} {\bibinfo {author} {\bibfnamefont {A.}~\bibnamefont
			{Osada}}, \bibinfo {author} {\bibfnamefont {Y.}~\bibnamefont {Ota}}, \bibinfo
		{author} {\bibfnamefont {R.}~\bibnamefont {Katsumi}}, \bibinfo {author}
		{\bibfnamefont {M.}~\bibnamefont {Kakuda}}, \bibinfo {author} {\bibfnamefont
			{S.}~\bibnamefont {Iwamoto}},\ and\ \bibinfo {author} {\bibfnamefont
			{Y.}~\bibnamefont {Arakawa}},\ }\bibfield  {title} {\bibinfo {title}
		{{Strongly coupled single-quantum-dot--cavity system integrated on a
				CMOS-processed silicon photonic chip}},\ }\href@noop {} {\bibfield  {journal}
		{\bibinfo  {journal} {Phys. Rev. Applied}\ }\textbf {\bibinfo {volume}
			{11}},\ \bibinfo {pages} {024071} (\bibinfo {year} {2019})}\BibitemShut
	{NoStop}%
	\bibitem [{\citenamefont {Zadeh}\ \emph {et~al.}(2016)\citenamefont {Zadeh},
		\citenamefont {Elshaari}, \citenamefont {J{\"o}ns}, \citenamefont {Fognini},
		\citenamefont {Dalacu}, \citenamefont {Poole}, \citenamefont {Reimer},\ and\
		\citenamefont {Zwiller}}]{Zadeh16}%
	\BibitemOpen
	\bibfield  {author} {\bibinfo {author} {\bibfnamefont {I.~E.}\ \bibnamefont
			{Zadeh}}, \bibinfo {author} {\bibfnamefont {A.~W.}\ \bibnamefont {Elshaari}},
		\bibinfo {author} {\bibfnamefont {K.~D.}\ \bibnamefont {J{\"o}ns}}, \bibinfo
		{author} {\bibfnamefont {A.}~\bibnamefont {Fognini}}, \bibinfo {author}
		{\bibfnamefont {D.}~\bibnamefont {Dalacu}}, \bibinfo {author} {\bibfnamefont
			{P.~J.}\ \bibnamefont {Poole}}, \bibinfo {author} {\bibfnamefont {M.~E.}\
			\bibnamefont {Reimer}},\ and\ \bibinfo {author} {\bibfnamefont
			{V.}~\bibnamefont {Zwiller}},\ }\bibfield  {title} {\bibinfo {title}
		{Deterministic integration of single photon sources in silicon based photonic
			circuits},\ }\href@noop {} {\bibfield  {journal} {\bibinfo  {journal} {Nano
				Letters}\ }\textbf {\bibinfo {volume} {16}},\ \bibinfo {pages} {2289}
		(\bibinfo {year} {2016})}\BibitemShut {NoStop}%
	\bibitem [{\citenamefont {Kim}\ \emph {et~al.}(2017)\citenamefont {Kim},
		\citenamefont {Aghaeimeibodi}, \citenamefont {Richardson}, \citenamefont
		{Leavitt}, \citenamefont {Englund},\ and\ \citenamefont {Waks}}]{Kim17}%
	\BibitemOpen
	\bibfield  {author} {\bibinfo {author} {\bibfnamefont {J.-H.}\ \bibnamefont
			{Kim}}, \bibinfo {author} {\bibfnamefont {S.}~\bibnamefont {Aghaeimeibodi}},
		\bibinfo {author} {\bibfnamefont {C.~J.}\ \bibnamefont {Richardson}},
		\bibinfo {author} {\bibfnamefont {R.~P.}\ \bibnamefont {Leavitt}}, \bibinfo
		{author} {\bibfnamefont {D.}~\bibnamefont {Englund}},\ and\ \bibinfo {author}
		{\bibfnamefont {E.}~\bibnamefont {Waks}},\ }\bibfield  {title} {\bibinfo
		{title} {Hybrid integration of solid-state quantum emitters on a silicon
			photonic chip},\ }\href@noop {} {\bibfield  {journal} {\bibinfo  {journal}
			{Nano Letters}\ }\textbf {\bibinfo {volume} {17}},\ \bibinfo {pages} {7394}
		(\bibinfo {year} {2017})}\BibitemShut {NoStop}%
	\bibitem [{\citenamefont {Aharonovich}\ \emph {et~al.}(2016)\citenamefont
		{Aharonovich}, \citenamefont {Englund},\ and\ \citenamefont
		{Toth}}]{Aharonovich16}%
	\BibitemOpen
	\bibfield  {author} {\bibinfo {author} {\bibfnamefont {I.}~\bibnamefont
			{Aharonovich}}, \bibinfo {author} {\bibfnamefont {D.}~\bibnamefont
			{Englund}},\ and\ \bibinfo {author} {\bibfnamefont {M.}~\bibnamefont
			{Toth}},\ }\bibfield  {title} {\bibinfo {title} {Solid-state single-photon
			emitters},\ }\href@noop {} {\bibfield  {journal} {\bibinfo  {journal} {Nature
				Photonics}\ }\textbf {\bibinfo {volume} {10}},\ \bibinfo {pages} {631}
		(\bibinfo {year} {2016})}\BibitemShut {NoStop}%
	\bibitem [{\citenamefont {Belhassen}\ \emph {et~al.}(2018)\citenamefont
		{Belhassen}, \citenamefont {Baboux}, \citenamefont {Yao}, \citenamefont
		{Amanti}, \citenamefont {Favero}, \citenamefont {Lema{\^\i}tre},
		\citenamefont {Kolthammer}, \citenamefont {Walmsley},\ and\ \citenamefont
		{Ducci}}]{Belhassen18}%
	\BibitemOpen
	\bibfield  {author} {\bibinfo {author} {\bibfnamefont {J.}~\bibnamefont
			{Belhassen}}, \bibinfo {author} {\bibfnamefont {F.}~\bibnamefont {Baboux}},
		\bibinfo {author} {\bibfnamefont {Q.}~\bibnamefont {Yao}}, \bibinfo {author}
		{\bibfnamefont {M.}~\bibnamefont {Amanti}}, \bibinfo {author} {\bibfnamefont
			{I.}~\bibnamefont {Favero}}, \bibinfo {author} {\bibfnamefont
			{A.}~\bibnamefont {Lema{\^\i}tre}}, \bibinfo {author} {\bibfnamefont
			{W.}~\bibnamefont {Kolthammer}}, \bibinfo {author} {\bibfnamefont
			{I.}~\bibnamefont {Walmsley}},\ and\ \bibinfo {author} {\bibfnamefont
			{S.}~\bibnamefont {Ducci}},\ }\bibfield  {title} {\bibinfo {title} {On-chip
			{III-V} monolithic integration of heralded single photon sources and
			beamsplitters},\ }\href@noop {} {\bibfield  {journal} {\bibinfo  {journal}
			{Applied Physics Letters}\ }\textbf {\bibinfo {volume} {112}},\ \bibinfo
		{pages} {071105} (\bibinfo {year} {2018})}\BibitemShut {NoStop}%
	\bibitem [{\citenamefont {Chen}\ \emph {et~al.}(2018)\citenamefont {Chen},
		\citenamefont {Auchter}, \citenamefont {Prilm{\"u}ller}, \citenamefont
		{Schlager}, \citenamefont {Kauten}, \citenamefont {Laiho}, \citenamefont
		{Pressl}, \citenamefont {Suchomel}, \citenamefont {Kamp}, \citenamefont
		{H{\"o}fling} \emph {et~al.}}]{Chen18}%
	\BibitemOpen
	\bibfield  {author} {\bibinfo {author} {\bibfnamefont {H.}~\bibnamefont
			{Chen}}, \bibinfo {author} {\bibfnamefont {S.}~\bibnamefont {Auchter}},
		\bibinfo {author} {\bibfnamefont {M.}~\bibnamefont {Prilm{\"u}ller}},
		\bibinfo {author} {\bibfnamefont {A.}~\bibnamefont {Schlager}}, \bibinfo
		{author} {\bibfnamefont {T.}~\bibnamefont {Kauten}}, \bibinfo {author}
		{\bibfnamefont {K.}~\bibnamefont {Laiho}}, \bibinfo {author} {\bibfnamefont
			{B.}~\bibnamefont {Pressl}}, \bibinfo {author} {\bibfnamefont
			{H.}~\bibnamefont {Suchomel}}, \bibinfo {author} {\bibfnamefont
			{M.}~\bibnamefont {Kamp}}, \bibinfo {author} {\bibfnamefont {S.}~\bibnamefont
			{H{\"o}fling}}, \emph {et~al.},\ }\bibfield  {title} {\bibinfo {title}
		{Time-bin entangled photon pairs from {Bragg-reflection} waveguides},\
	}\href@noop {} {\bibfield  {journal} {\bibinfo  {journal} {APL Photonics}\
		}\textbf {\bibinfo {volume} {3}},\ \bibinfo {pages} {080804} (\bibinfo {year}
		{2018})}\BibitemShut {NoStop}%
	\bibitem [{\citenamefont {Wang}\ \emph {et~al.}(2021)\citenamefont {Wang},
		\citenamefont {J{\"o}ns},\ and\ \citenamefont {Sun}}]{Wang21}%
	\BibitemOpen
	\bibfield  {author} {\bibinfo {author} {\bibfnamefont {Y.}~\bibnamefont
			{Wang}}, \bibinfo {author} {\bibfnamefont {K.~D.}\ \bibnamefont {J{\"o}ns}},\
		and\ \bibinfo {author} {\bibfnamefont {Z.}~\bibnamefont {Sun}},\ }\bibfield
	{title} {\bibinfo {title} {Integrated photon-pair sources with nonlinear
			optics},\ }\href@noop {} {\bibfield  {journal} {\bibinfo  {journal} {Applied
				Physics Reviews}\ }\textbf {\bibinfo {volume} {8}} (\bibinfo {year}
		{2021})}\BibitemShut {NoStop}%
	\bibitem [{\citenamefont {Andersen}\ \emph {et~al.}(2016)\citenamefont
		{Andersen}, \citenamefont {Gehring}, \citenamefont {Marquardt},\ and\
		\citenamefont {Leuchs}}]{Andersen16}%
	\BibitemOpen
	\bibfield  {author} {\bibinfo {author} {\bibfnamefont {U.~L.}\ \bibnamefont
			{Andersen}}, \bibinfo {author} {\bibfnamefont {T.}~\bibnamefont {Gehring}},
		\bibinfo {author} {\bibfnamefont {C.}~\bibnamefont {Marquardt}},\ and\
		\bibinfo {author} {\bibfnamefont {G.}~\bibnamefont {Leuchs}},\ }\bibfield
	{title} {\bibinfo {title} {30 years of squeezed light generation},\
	}\href@noop {} {\bibfield  {journal} {\bibinfo  {journal} {Physica Scripta}\
		}\textbf {\bibinfo {volume} {91}},\ \bibinfo {pages} {053001} (\bibinfo
		{year} {2016})}\BibitemShut {NoStop}%
	\bibitem [{\citenamefont {Brodutch}\ \emph {et~al.}(2018)\citenamefont
		{Brodutch}, \citenamefont {Marchildon},\ and\ \citenamefont
		{Helmy}}]{Brodutch18}%
	\BibitemOpen
	\bibfield  {author} {\bibinfo {author} {\bibfnamefont {A.}~\bibnamefont
			{Brodutch}}, \bibinfo {author} {\bibfnamefont {R.}~\bibnamefont
			{Marchildon}},\ and\ \bibinfo {author} {\bibfnamefont {A.~S.}\ \bibnamefont
			{Helmy}},\ }\bibfield  {title} {\bibinfo {title} {Dynamically reconfigurable
			sources for arbitrary {Gaussian} states in integrated photonics circuits},\
	}\href@noop {} {\bibfield  {journal} {\bibinfo  {journal} {Optics Express}\
		}\textbf {\bibinfo {volume} {26}},\ \bibinfo {pages} {17635} (\bibinfo {year}
		{2018})}\BibitemShut {NoStop}%
	\bibitem [{\citenamefont {Mahmudlu}\ \emph {et~al.}(2023)\citenamefont
		{Mahmudlu}, \citenamefont {Johanning}, \citenamefont {van Rees},
		\citenamefont {Kashi}, \citenamefont {Epping}, \citenamefont {Haldar},
		\citenamefont {Boller},\ and\ \citenamefont {Kues}}]{kues23}%
	\BibitemOpen
	\bibfield  {author} {\bibinfo {author} {\bibfnamefont {H.}~\bibnamefont
			{Mahmudlu}}, \bibinfo {author} {\bibfnamefont {R.}~\bibnamefont {Johanning}},
		\bibinfo {author} {\bibfnamefont {A.}~\bibnamefont {van Rees}}, \bibinfo
		{author} {\bibfnamefont {A.~K.}\ \bibnamefont {Kashi}}, \bibinfo {author}
		{\bibfnamefont {J.~P.}\ \bibnamefont {Epping}}, \bibinfo {author}
		{\bibfnamefont {R.}~\bibnamefont {Haldar}}, \bibinfo {author} {\bibfnamefont
			{K.-J.}\ \bibnamefont {Boller}},\ and\ \bibinfo {author} {\bibfnamefont
			{M.}~\bibnamefont {Kues}},\ }\bibfield  {title} {\bibinfo {title} {Fully
			on-chip photonic turnkey quantum source for entangled qubit/qudit state
			generation},\ }\href@noop {} {\bibfield  {journal} {\bibinfo  {journal}
			{Nature Photonics}\ }\textbf {\bibinfo {volume} {17}},\ \bibinfo {pages}
		{518} (\bibinfo {year} {2023})}\BibitemShut {NoStop}%
	\bibitem [{\citenamefont {Yeh}\ and\ \citenamefont
		{Yariv}(1976)}]{BRW_Original}%
	\BibitemOpen
	\bibfield  {author} {\bibinfo {author} {\bibfnamefont {P.}~\bibnamefont
			{Yeh}}\ and\ \bibinfo {author} {\bibfnamefont {A.}~\bibnamefont {Yariv}},\
	}\bibfield  {title} {\bibinfo {title} {Bragg reflection waveguides},\
	}\href@noop {} {\bibfield  {journal} {\bibinfo  {journal} {Optics
				Commununications}\ }\textbf {\bibinfo {volume} {19}},\ \bibinfo {pages} {427}
		(\bibinfo {year} {1976})}\BibitemShut {NoStop}%
	\bibitem [{\citenamefont {Appas}\ \emph {et~al.}(2022)\citenamefont {Appas},
		\citenamefont {Meskine}, \citenamefont {Lema\^{i}tre}, \citenamefont
		{Morassi}, \citenamefont {Baboux}, \citenamefont {Amanti},\ and\
		\citenamefont {Ducci}}]{Review_Feli}%
	\BibitemOpen
	\bibfield  {author} {\bibinfo {author} {\bibfnamefont {F.}~\bibnamefont
			{Appas}}, \bibinfo {author} {\bibfnamefont {O.}~\bibnamefont {Meskine}},
		\bibinfo {author} {\bibfnamefont {A.}~\bibnamefont {Lema\^{i}tre}}, \bibinfo
		{author} {\bibfnamefont {M.}~\bibnamefont {Morassi}}, \bibinfo {author}
		{\bibfnamefont {F.}~\bibnamefont {Baboux}}, \bibinfo {author} {\bibfnamefont
			{M.~I.}\ \bibnamefont {Amanti}},\ and\ \bibinfo {author} {\bibfnamefont
			{S.}~\bibnamefont {Ducci}},\ }\bibfield  {title} {\bibinfo {title} {Nonlinear
			quantum photonics with algaas bragg-reflection waveguides},\ }\href@noop {}
	{\bibfield  {journal} {\bibinfo  {journal} {Journal of Lightwave Technology}\
		}\textbf {\bibinfo {volume} {40}},\ \bibinfo {pages} {7658} (\bibinfo {year}
		{2022})}\BibitemShut {NoStop}%
	\bibitem [{\citenamefont {Ansari}\ \emph {et~al.}(2018)\citenamefont {Ansari},
		\citenamefont {Donohue}, \citenamefont {Brecht},\ and\ \citenamefont
		{Silberhorn}}]{Ansari18b}%
	\BibitemOpen
	\bibfield  {author} {\bibinfo {author} {\bibfnamefont {V.}~\bibnamefont
			{Ansari}}, \bibinfo {author} {\bibfnamefont {J.~M.}\ \bibnamefont {Donohue}},
		\bibinfo {author} {\bibfnamefont {B.}~\bibnamefont {Brecht}},\ and\ \bibinfo
		{author} {\bibfnamefont {C.}~\bibnamefont {Silberhorn}},\ }\bibfield  {title}
	{\bibinfo {title} {Tailoring nonlinear processes for quantum optics with
			pulsed temporal-mode encodings},\ }\href@noop {} {\bibfield  {journal}
		{\bibinfo  {journal} {Optica}\ }\textbf {\bibinfo {volume} {5}},\ \bibinfo
		{pages} {534} (\bibinfo {year} {2018})}\BibitemShut {NoStop}%
	\bibitem [{\citenamefont {Gianani}\ \emph {et~al.}(2020)\citenamefont
		{Gianani}, \citenamefont {Sbroscia},\ and\ \citenamefont
		{Barbieri}}]{Gianani20}%
	\BibitemOpen
	\bibfield  {author} {\bibinfo {author} {\bibfnamefont {I.}~\bibnamefont
			{Gianani}}, \bibinfo {author} {\bibfnamefont {M.}~\bibnamefont {Sbroscia}},\
		and\ \bibinfo {author} {\bibfnamefont {M.}~\bibnamefont {Barbieri}},\
	}\bibfield  {title} {\bibinfo {title} {Measuring the time--frequency
			properties of photon pairs: A short review},\ }\href@noop {} {\bibfield
		{journal} {\bibinfo  {journal} {AVS Quantum Science}\ }\textbf {\bibinfo
			{volume} {2}} (\bibinfo {year} {2020})}\BibitemShut {NoStop}%
	\bibitem [{\citenamefont {Maltese}\ \emph
		{et~al.}(2020{\natexlab{a}})\citenamefont {Maltese}, \citenamefont {Amanti},
		\citenamefont {Appas}, \citenamefont {Sinnl}, \citenamefont {Lema{\^\i}tre},
		\citenamefont {Milman}, \citenamefont {Baboux},\ and\ \citenamefont
		{Ducci}}]{Maltese20}%
	\BibitemOpen
	\bibfield  {author} {\bibinfo {author} {\bibfnamefont {G.}~\bibnamefont
			{Maltese}}, \bibinfo {author} {\bibfnamefont {M.}~\bibnamefont {Amanti}},
		\bibinfo {author} {\bibfnamefont {F.}~\bibnamefont {Appas}}, \bibinfo
		{author} {\bibfnamefont {G.}~\bibnamefont {Sinnl}}, \bibinfo {author}
		{\bibfnamefont {A.}~\bibnamefont {Lema{\^\i}tre}}, \bibinfo {author}
		{\bibfnamefont {P.}~\bibnamefont {Milman}}, \bibinfo {author} {\bibfnamefont
			{F.}~\bibnamefont {Baboux}},\ and\ \bibinfo {author} {\bibfnamefont
			{S.}~\bibnamefont {Ducci}},\ }\bibfield  {title} {\bibinfo {title}
		{Generation and symmetry control of quantum frequency combs},\ }\href@noop {}
	{\bibfield  {journal} {\bibinfo  {journal} {npj Quantum Information}\
		}\textbf {\bibinfo {volume} {6}},\ \bibinfo {pages} {13} (\bibinfo {year}
		{2020}{\natexlab{a}})}\BibitemShut {NoStop}%
	\bibitem [{\citenamefont {Appas}\ \emph {et~al.}(2021)\citenamefont {Appas},
		\citenamefont {Baboux}, \citenamefont {Amanti}, \citenamefont
		{Lema{\^\i}tre}, \citenamefont {Boitier}, \citenamefont {Diamanti},\ and\
		\citenamefont {Ducci}}]{Appas21}%
	\BibitemOpen
	\bibfield  {author} {\bibinfo {author} {\bibfnamefont {F.}~\bibnamefont
			{Appas}}, \bibinfo {author} {\bibfnamefont {F.}~\bibnamefont {Baboux}},
		\bibinfo {author} {\bibfnamefont {M.}~\bibnamefont {Amanti}}, \bibinfo
		{author} {\bibfnamefont {A.}~\bibnamefont {Lema{\^\i}tre}}, \bibinfo {author}
		{\bibfnamefont {F.}~\bibnamefont {Boitier}}, \bibinfo {author} {\bibfnamefont
			{E.}~\bibnamefont {Diamanti}},\ and\ \bibinfo {author} {\bibfnamefont
			{S.}~\bibnamefont {Ducci}},\ }\bibfield  {title} {\bibinfo {title} {{Flexible
				entanglement-distribution network with an AlGaAs chip for secure
				communications}},\ }\href@noop {} {\bibfield  {journal} {\bibinfo  {journal}
			{npj Quantum Information}\ }\textbf {\bibinfo {volume} {7}},\ \bibinfo
		{pages} {1} (\bibinfo {year} {2021})}\BibitemShut {NoStop}%
	\bibitem [{\citenamefont {Bücher}\ \emph {et~al.}(1994)\citenamefont
		{Bücher}, \citenamefont {Bruns},\ and\ \citenamefont
		{Wagemann}}]{Silicon_absCoeff}%
	\BibitemOpen
	\bibfield  {author} {\bibinfo {author} {\bibfnamefont {K.}~\bibnamefont
			{Bücher}}, \bibinfo {author} {\bibfnamefont {J.}~\bibnamefont {Bruns}},\
		and\ \bibinfo {author} {\bibfnamefont {H.~G.}\ \bibnamefont {Wagemann}},\
	}\bibfield  {title} {\bibinfo {title} {Absorption coefficient of silicon: An
			assessment of measurements and the simulation of temperature variation},\
	}\href@noop {} {\bibfield  {journal} {\bibinfo  {journal} {Journal of Applied
				Physics}\ }\textbf {\bibinfo {volume} {75}},\ \bibinfo {pages} {1127}
		(\bibinfo {year} {1994})}\BibitemShut {NoStop}%
	\bibitem [{\citenamefont {Sun}\ \emph {et~al.}(2009)\citenamefont {Sun},
		\citenamefont {Liu},\ and\ \citenamefont {Yariv}}]{Sun09}%
	\BibitemOpen
	\bibfield  {author} {\bibinfo {author} {\bibfnamefont {X.}~\bibnamefont
			{Sun}}, \bibinfo {author} {\bibfnamefont {H.-C.}\ \bibnamefont {Liu}},\ and\
		\bibinfo {author} {\bibfnamefont {A.}~\bibnamefont {Yariv}},\ }\bibfield
	{title} {\bibinfo {title} {Adiabaticity criterion and the shortest adiabatic
			mode transformer in a coupled-waveguide system},\ }\href@noop {} {\bibfield
		{journal} {\bibinfo  {journal} {Optics Letters}\ }\textbf {\bibinfo {volume}
			{34}},\ \bibinfo {pages} {280} (\bibinfo {year} {2009})}\BibitemShut
	{NoStop}%
	\bibitem [{\citenamefont {Fu}\ \emph {et~al.}(2014)\citenamefont {Fu},
		\citenamefont {Ye}, \citenamefont {Tang},\ and\ \citenamefont {Chu}}]{Fu14}%
	\BibitemOpen
	\bibfield  {author} {\bibinfo {author} {\bibfnamefont {Y.}~\bibnamefont
			{Fu}}, \bibinfo {author} {\bibfnamefont {T.}~\bibnamefont {Ye}}, \bibinfo
		{author} {\bibfnamefont {W.}~\bibnamefont {Tang}},\ and\ \bibinfo {author}
		{\bibfnamefont {T.}~\bibnamefont {Chu}},\ }\bibfield  {title} {\bibinfo
		{title} {Efficient adiabatic silicon-on-insulator waveguide taper},\
	}\href@noop {} {\bibfield  {journal} {\bibinfo  {journal} {Photonics
				Research}\ }\textbf {\bibinfo {volume} {2}},\ \bibinfo {pages} {A41}
		(\bibinfo {year} {2014})}\BibitemShut {NoStop}%
	\bibitem [{\citenamefont {Keyvaninia}\ \emph {et~al.}(2013)\citenamefont
		{Keyvaninia}, \citenamefont {Roelkens}, \citenamefont {Van~Thourhout},
		\citenamefont {Jany}, \citenamefont {Lamponi}, \citenamefont {Le~Liepvre},
		\citenamefont {Lelarge}, \citenamefont {Make}, \citenamefont {Duan},
		\citenamefont {Bordel},\ and\ \citenamefont {Fedeli}}]{MolecularBonding}%
	\BibitemOpen
	\bibfield  {author} {\bibinfo {author} {\bibfnamefont {S.}~\bibnamefont
			{Keyvaninia}}, \bibinfo {author} {\bibfnamefont {G.}~\bibnamefont
			{Roelkens}}, \bibinfo {author} {\bibfnamefont {D.}~\bibnamefont
			{Van~Thourhout}}, \bibinfo {author} {\bibfnamefont {C.}~\bibnamefont {Jany}},
		\bibinfo {author} {\bibfnamefont {M.}~\bibnamefont {Lamponi}}, \bibinfo
		{author} {\bibfnamefont {A.}~\bibnamefont {Le~Liepvre}}, \bibinfo {author}
		{\bibfnamefont {F.}~\bibnamefont {Lelarge}}, \bibinfo {author} {\bibfnamefont
			{D.}~\bibnamefont {Make}}, \bibinfo {author} {\bibfnamefont {G.}~\bibnamefont
			{Duan}}, \bibinfo {author} {\bibfnamefont {D.}~\bibnamefont {Bordel}},\ and\
		\bibinfo {author} {\bibfnamefont {J.}~\bibnamefont {Fedeli}},\ }\bibfield
	{title} {\bibinfo {title} {Demonstration of a heterogeneously integrated
			iii-v/soi single wavelength tunable laser},\ }\href@noop {} {\bibfield
		{journal} {\bibinfo  {journal} {Opt. Express}\ }\textbf {\bibinfo {volume}
			{21}},\ \bibinfo {pages} {3784} (\bibinfo {year} {2013})}\BibitemShut
	{NoStop}%
	\bibitem [{\citenamefont {Tang}\ \emph {et~al.}(2018)\citenamefont {Tang},
		\citenamefont {Chen}, \citenamefont {Kho}, \citenamefont {Hsieh},
		\citenamefont {Chang},\ and\ \citenamefont {Chen}}]{Tang2018InvestigationAO}%
	\BibitemOpen
	\bibfield  {author} {\bibinfo {author} {\bibfnamefont {Y.-S.}\ \bibnamefont
			{Tang}}, \bibinfo {author} {\bibfnamefont {H.-C.}\ \bibnamefont {Chen}},
		\bibinfo {author} {\bibfnamefont {Y.~T.}\ \bibnamefont {Kho}}, \bibinfo
		{author} {\bibfnamefont {Y.-S.}\ \bibnamefont {Hsieh}}, \bibinfo {author}
		{\bibfnamefont {Y.-J.}\ \bibnamefont {Chang}},\ and\ \bibinfo {author}
		{\bibfnamefont {K.-N.}\ \bibnamefont {Chen}},\ }\bibfield  {title} {\bibinfo
		{title} {Investigation and optimization of ultrathin buffer layers used in
			{Cu/Sn} eutectic bonding},\ }\href@noop {} {\bibfield  {journal} {\bibinfo
			{journal} {IEEE Transactions on Components, Packaging and Manufacturing
				Technology}\ }\textbf {\bibinfo {volume} {8}},\ \bibinfo {pages} {1225}
		(\bibinfo {year} {2018})}\BibitemShut {NoStop}%
	\bibitem [{\citenamefont {Alaskar}\ \emph {et~al.}(2014)\citenamefont
		{Alaskar}, \citenamefont {Arafin}, \citenamefont {Wickramaratne},
		\citenamefont {Zurbuchen}, \citenamefont {He}, \citenamefont {McKay},
		\citenamefont {Lin}, \citenamefont {Goorsky}, \citenamefont {Lake},\ and\
		\citenamefont {Wang}}]{DirectRegrowth}%
	\BibitemOpen
	\bibfield  {author} {\bibinfo {author} {\bibfnamefont {Y.}~\bibnamefont
			{Alaskar}}, \bibinfo {author} {\bibfnamefont {S.}~\bibnamefont {Arafin}},
		\bibinfo {author} {\bibfnamefont {D.}~\bibnamefont {Wickramaratne}}, \bibinfo
		{author} {\bibfnamefont {M.}~\bibnamefont {Zurbuchen}}, \bibinfo {author}
		{\bibfnamefont {L.}~\bibnamefont {He}}, \bibinfo {author} {\bibfnamefont
			{J.}~\bibnamefont {McKay}}, \bibinfo {author} {\bibfnamefont
			{Q.}~\bibnamefont {Lin}}, \bibinfo {author} {\bibfnamefont {M.}~\bibnamefont
			{Goorsky}}, \bibinfo {author} {\bibfnamefont {R.}~\bibnamefont {Lake}},\ and\
		\bibinfo {author} {\bibfnamefont {K.}~\bibnamefont {Wang}},\ }\bibfield
	{title} {\bibinfo {title} {Towards van der waals epitaxial growth of {GaAs on
				Si} using a graphene buffer layer},\ }\href@noop {} {\bibfield  {journal}
		{\bibinfo  {journal} {T. Adv. Funct. Mater.}\ }\textbf {\bibinfo {volume}
			{24}},\ \bibinfo {pages} {6629} (\bibinfo {year} {2014})}\BibitemShut
	{NoStop}%
	\bibitem [{\citenamefont {Haq}\ \emph {et~al.}(2020)\citenamefont {Haq},
		\citenamefont {Kumari}, \citenamefont {Van}, \citenamefont {Zhang},
		\citenamefont {Gocalinska}, \citenamefont {Pelucchi}, \citenamefont
		{Corbett},\ and\ \citenamefont {Roelkens}}]{MicroTransfertPrint}%
	\BibitemOpen
	\bibfield  {author} {\bibinfo {author} {\bibfnamefont {B.}~\bibnamefont
			{Haq}}, \bibinfo {author} {\bibfnamefont {S.}~\bibnamefont {Kumari}},
		\bibinfo {author} {\bibfnamefont {K.}~\bibnamefont {Van}}, \bibinfo {author}
		{\bibfnamefont {J.}~\bibnamefont {Zhang}}, \bibinfo {author} {\bibfnamefont
			{A.}~\bibnamefont {Gocalinska}}, \bibinfo {author} {\bibfnamefont
			{E.}~\bibnamefont {Pelucchi}}, \bibinfo {author} {\bibfnamefont
			{B.}~\bibnamefont {Corbett}},\ and\ \bibinfo {author} {\bibfnamefont
			{G.}~\bibnamefont {Roelkens}},\ }\bibfield  {title} {\bibinfo {title}
		{Micro-transfer-printed {III-V-on-Silicon C-band} semiconductor optical
			amplifiers},\ }\href@noop {} {\bibfield  {journal} {\bibinfo  {journal}
			{Laser \& Photonics Reviews}\ }\textbf {\bibinfo {volume} {14}},\ \bibinfo
		{pages} {1900364} (\bibinfo {year} {2020})}\BibitemShut {NoStop}%
	\bibitem [{\citenamefont {Crosnier}\ \emph {et~al.}(2017)\citenamefont
		{Crosnier}, \citenamefont {Sanchez}, \citenamefont {Bouchoule}, \citenamefont
		{Monnier}, \citenamefont {Beaudoin}, \citenamefont {Sagnes}, \citenamefont
		{Raj},\ and\ \citenamefont {Raineri}}]{Crosnier17}%
	\BibitemOpen
	\bibfield  {author} {\bibinfo {author} {\bibfnamefont {G.}~\bibnamefont
			{Crosnier}}, \bibinfo {author} {\bibfnamefont {D.}~\bibnamefont {Sanchez}},
		\bibinfo {author} {\bibfnamefont {S.}~\bibnamefont {Bouchoule}}, \bibinfo
		{author} {\bibfnamefont {P.}~\bibnamefont {Monnier}}, \bibinfo {author}
		{\bibfnamefont {G.}~\bibnamefont {Beaudoin}}, \bibinfo {author}
		{\bibfnamefont {I.}~\bibnamefont {Sagnes}}, \bibinfo {author} {\bibfnamefont
			{R.}~\bibnamefont {Raj}},\ and\ \bibinfo {author} {\bibfnamefont
			{F.}~\bibnamefont {Raineri}},\ }\bibfield  {title} {\bibinfo {title} {Hybrid
			indium phosphide-on-silicon nanolaser diode},\ }\href@noop {} {\bibfield
		{journal} {\bibinfo  {journal} {Nature Photonics}\ }\textbf {\bibinfo
			{volume} {11}},\ \bibinfo {pages} {297} (\bibinfo {year} {2017})}\BibitemShut
	{NoStop}%
	\bibitem [{\citenamefont {Atzeni}\ \emph {et~al.}(2018)\citenamefont {Atzeni},
		\citenamefont {Rab}, \citenamefont {Corrielli}, \citenamefont {Polino},
		\citenamefont {Valeri}, \citenamefont {Mataloni}, \citenamefont {Spagnolo},
		\citenamefont {Crespi}, \citenamefont {Sciarrino},\ and\ \citenamefont
		{Osellame}}]{Atzeni18}%
	\BibitemOpen
	\bibfield  {author} {\bibinfo {author} {\bibfnamefont {S.}~\bibnamefont
			{Atzeni}}, \bibinfo {author} {\bibfnamefont {A.~S.}\ \bibnamefont {Rab}},
		\bibinfo {author} {\bibfnamefont {G.}~\bibnamefont {Corrielli}}, \bibinfo
		{author} {\bibfnamefont {E.}~\bibnamefont {Polino}}, \bibinfo {author}
		{\bibfnamefont {M.}~\bibnamefont {Valeri}}, \bibinfo {author} {\bibfnamefont
			{P.}~\bibnamefont {Mataloni}}, \bibinfo {author} {\bibfnamefont
			{N.}~\bibnamefont {Spagnolo}}, \bibinfo {author} {\bibfnamefont
			{A.}~\bibnamefont {Crespi}}, \bibinfo {author} {\bibfnamefont
			{F.}~\bibnamefont {Sciarrino}},\ and\ \bibinfo {author} {\bibfnamefont
			{R.}~\bibnamefont {Osellame}},\ }\bibfield  {title} {\bibinfo {title}
		{Integrated sources of entangled photons at the telecom wavelength in
			femtosecond-laser-written circuits},\ }\href@noop {} {\bibfield  {journal}
		{\bibinfo  {journal} {Optica}\ }\textbf {\bibinfo {volume} {5}},\ \bibinfo
		{pages} {311} (\bibinfo {year} {2018})}\BibitemShut {NoStop}%
	\bibitem [{\citenamefont {Sabattoli}\ \emph {et~al.}(2022)\citenamefont
		{Sabattoli}, \citenamefont {Gianini}, \citenamefont {Simbula}, \citenamefont
		{Clementi}, \citenamefont {Fincato}, \citenamefont {Boeuf}, \citenamefont
		{Liscidini}, \citenamefont {Galli},\ and\ \citenamefont
		{Bajoni}}]{Sabattoli22}%
	\BibitemOpen
	\bibfield  {author} {\bibinfo {author} {\bibfnamefont {F.~A.}\ \bibnamefont
			{Sabattoli}}, \bibinfo {author} {\bibfnamefont {L.}~\bibnamefont {Gianini}},
		\bibinfo {author} {\bibfnamefont {A.}~\bibnamefont {Simbula}}, \bibinfo
		{author} {\bibfnamefont {M.}~\bibnamefont {Clementi}}, \bibinfo {author}
		{\bibfnamefont {A.}~\bibnamefont {Fincato}}, \bibinfo {author} {\bibfnamefont
			{F.}~\bibnamefont {Boeuf}}, \bibinfo {author} {\bibfnamefont
			{M.}~\bibnamefont {Liscidini}}, \bibinfo {author} {\bibfnamefont
			{M.}~\bibnamefont {Galli}},\ and\ \bibinfo {author} {\bibfnamefont
			{D.}~\bibnamefont {Bajoni}},\ }\bibfield  {title} {\bibinfo {title} {A
			silicon source of frequency-bin entangled photons},\ }\href@noop {}
	{\bibfield  {journal} {\bibinfo  {journal} {Opt. Lett.}\ }\textbf {\bibinfo
			{volume} {47}},\ \bibinfo {pages} {6201} (\bibinfo {year}
		{2022})}\BibitemShut {NoStop}%
	\bibitem [{\citenamefont {Oser}\ \emph {et~al.}(2020)\citenamefont {Oser},
		\citenamefont {Tanzilli}, \citenamefont {Mazeas}, \citenamefont
		{Alonso-Ramos}, \citenamefont {Le~Roux}, \citenamefont {Sauder},
		\citenamefont {Hua}, \citenamefont {Alibart}, \citenamefont {Vivien},
		\citenamefont {Cassan} \emph {et~al.}}]{Oser20}%
	\BibitemOpen
	\bibfield  {author} {\bibinfo {author} {\bibfnamefont {D.}~\bibnamefont
			{Oser}}, \bibinfo {author} {\bibfnamefont {S.}~\bibnamefont {Tanzilli}},
		\bibinfo {author} {\bibfnamefont {F.}~\bibnamefont {Mazeas}}, \bibinfo
		{author} {\bibfnamefont {C.}~\bibnamefont {Alonso-Ramos}}, \bibinfo {author}
		{\bibfnamefont {X.}~\bibnamefont {Le~Roux}}, \bibinfo {author} {\bibfnamefont
			{G.}~\bibnamefont {Sauder}}, \bibinfo {author} {\bibfnamefont
			{X.}~\bibnamefont {Hua}}, \bibinfo {author} {\bibfnamefont {O.}~\bibnamefont
			{Alibart}}, \bibinfo {author} {\bibfnamefont {L.}~\bibnamefont {Vivien}},
		\bibinfo {author} {\bibfnamefont {{\'E}.}~\bibnamefont {Cassan}}, \emph
		{et~al.},\ }\bibfield  {title} {\bibinfo {title} {High-quality photonic
			entanglement out of a stand-alone silicon chip},\ }\href@noop {} {\bibfield
		{journal} {\bibinfo  {journal} {npj Quantum Information}\ }\textbf {\bibinfo
			{volume} {6}},\ \bibinfo {pages} {31} (\bibinfo {year} {2020})}\BibitemShut
	{NoStop}%
	\bibitem [{\citenamefont {Maltese}\ \emph
		{et~al.}(2020{\natexlab{b}})\citenamefont {Maltese}, \citenamefont {Amanti},
		\citenamefont {Appas}, \citenamefont {Sinnl}, \citenamefont {Lemaître},
		\citenamefont {Milman}, \citenamefont {Baboux},\ and\ \citenamefont
		{Ducci}}]{giorgio20}%
	\BibitemOpen
	\bibfield  {author} {\bibinfo {author} {\bibfnamefont {G.}~\bibnamefont
			{Maltese}}, \bibinfo {author} {\bibfnamefont {M.~I.}\ \bibnamefont {Amanti}},
		\bibinfo {author} {\bibfnamefont {F.}~\bibnamefont {Appas}}, \bibinfo
		{author} {\bibfnamefont {G.}~\bibnamefont {Sinnl}}, \bibinfo {author}
		{\bibfnamefont {A.}~\bibnamefont {Lemaître}}, \bibinfo {author}
		{\bibfnamefont {P.}~\bibnamefont {Milman}}, \bibinfo {author} {\bibfnamefont
			{F.}~\bibnamefont {Baboux}},\ and\ \bibinfo {author} {\bibfnamefont
			{S.}~\bibnamefont {Ducci}},\ }\bibfield  {title} {\bibinfo {title}
		{Generation and symmetry control of quantum frequency combs},\ }\href@noop {}
	{\bibfield  {journal} {\bibinfo  {journal} {npj Quantum Information}\
		}\textbf {\bibinfo {volume} {6}} (\bibinfo {year}
		{2020}{\natexlab{b}})}\BibitemShut {NoStop}%
	\bibitem [{\citenamefont {Kaiser}\ \emph {et~al.}(2018)\citenamefont {Kaiser},
		\citenamefont {Vergyris}, \citenamefont {Aktas}, \citenamefont {Babin},
		\citenamefont {Labont{\'e}},\ and\ \citenamefont {Tanzilli}}]{Kaiser18}%
	\BibitemOpen
	\bibfield  {author} {\bibinfo {author} {\bibfnamefont {F.}~\bibnamefont
			{Kaiser}}, \bibinfo {author} {\bibfnamefont {P.}~\bibnamefont {Vergyris}},
		\bibinfo {author} {\bibfnamefont {D.}~\bibnamefont {Aktas}}, \bibinfo
		{author} {\bibfnamefont {C.}~\bibnamefont {Babin}}, \bibinfo {author}
		{\bibfnamefont {L.}~\bibnamefont {Labont{\'e}}},\ and\ \bibinfo {author}
		{\bibfnamefont {S.}~\bibnamefont {Tanzilli}},\ }\bibfield  {title} {\bibinfo
		{title} {Quantum enhancement of accuracy and precision in optical
			interferometry},\ }\href@noop {} {\bibfield  {journal} {\bibinfo  {journal}
			{Light: Science \& Applications}\ }\textbf {\bibinfo {volume} {7}},\ \bibinfo
		{pages} {17163} (\bibinfo {year} {2018})}\BibitemShut {NoStop}%
	\bibitem [{\citenamefont {Franson}(1989)}]{Franson_Original}%
	\BibitemOpen
	\bibfield  {author} {\bibinfo {author} {\bibfnamefont {J.~D.}\ \bibnamefont
			{Franson}},\ }\bibfield  {title} {\bibinfo {title} {Bell inequality for
			position and time},\ }\href@noop {} {\bibfield  {journal} {\bibinfo
			{journal} {Physical Review Letters}\ }\textbf {\bibinfo {volume} {62}},\
		\bibinfo {pages} {2205} (\bibinfo {year} {1989})}\BibitemShut {NoStop}%
	\bibitem [{\citenamefont {Thew}\ \emph {et~al.}(2002)\citenamefont {Thew},
		\citenamefont {Tanzilli}, \citenamefont {Tittel}, \citenamefont {Zbinden},\
		and\ \citenamefont {Gisin}}]{Folded_Franson}%
	\BibitemOpen
	\bibfield  {author} {\bibinfo {author} {\bibfnamefont {R.~T.}\ \bibnamefont
			{Thew}}, \bibinfo {author} {\bibfnamefont {S.}~\bibnamefont {Tanzilli}},
		\bibinfo {author} {\bibfnamefont {W.}~\bibnamefont {Tittel}}, \bibinfo
		{author} {\bibfnamefont {H.}~\bibnamefont {Zbinden}},\ and\ \bibinfo {author}
		{\bibfnamefont {N.}~\bibnamefont {Gisin}},\ }\bibfield  {title} {\bibinfo
		{title} {Experimental investigation of the robustness of partially entangled
			qubits over 11 km},\ }\href@noop {} {\bibfield  {journal} {\bibinfo
			{journal} {Physical Review A}\ }\textbf {\bibinfo {volume} {66}},\ \bibinfo
		{pages} {062304} (\bibinfo {year} {2002})}\BibitemShut {NoStop}%
	\bibitem [{\citenamefont {Tittel}\ \emph {et~al.}(1999)\citenamefont {Tittel},
		\citenamefont {Brendel}, \citenamefont {Gisin},\ and\ \citenamefont
		{Zbinden}}]{Zbinden_Bell}%
	\BibitemOpen
	\bibfield  {author} {\bibinfo {author} {\bibfnamefont {W.}~\bibnamefont
			{Tittel}}, \bibinfo {author} {\bibfnamefont {J.}~\bibnamefont {Brendel}},
		\bibinfo {author} {\bibfnamefont {N.}~\bibnamefont {Gisin}},\ and\ \bibinfo
		{author} {\bibfnamefont {H.}~\bibnamefont {Zbinden}},\ }\bibfield  {title}
	{\bibinfo {title} {Long-distance bell-type tests using energy-time entangled
			photons},\ }\href@noop {} {\bibfield  {journal} {\bibinfo  {journal}
			{Physical Review A}\ }\textbf {\bibinfo {volume} {59}},\ \bibinfo {pages}
		{4150} (\bibinfo {year} {1999})}\BibitemShut {NoStop}%
	\bibitem [{\citenamefont {Szelag}\ \emph {et~al.}(2019)\citenamefont {Szelag},
		\citenamefont {Hassan}, \citenamefont {Adelmini}, \citenamefont {Ghegin},
		\citenamefont {Rodriguez}, \citenamefont {Nemouchi}, \citenamefont
		{Brianceau}, \citenamefont {Vermande}, \citenamefont {Schembri},
		\citenamefont {Carrara} \emph {et~al.}}]{Szelag19}%
	\BibitemOpen
	\bibfield  {author} {\bibinfo {author} {\bibfnamefont {B.}~\bibnamefont
			{Szelag}}, \bibinfo {author} {\bibfnamefont {K.}~\bibnamefont {Hassan}},
		\bibinfo {author} {\bibfnamefont {L.}~\bibnamefont {Adelmini}}, \bibinfo
		{author} {\bibfnamefont {E.}~\bibnamefont {Ghegin}}, \bibinfo {author}
		{\bibfnamefont {P.}~\bibnamefont {Rodriguez}}, \bibinfo {author}
		{\bibfnamefont {F.}~\bibnamefont {Nemouchi}}, \bibinfo {author}
		{\bibfnamefont {P.}~\bibnamefont {Brianceau}}, \bibinfo {author}
		{\bibfnamefont {E.}~\bibnamefont {Vermande}}, \bibinfo {author}
		{\bibfnamefont {A.}~\bibnamefont {Schembri}}, \bibinfo {author}
		{\bibfnamefont {D.}~\bibnamefont {Carrara}}, \emph {et~al.},\ }\bibfield
	{title} {\bibinfo {title} {{Hybrid III--V/silicon technology for laser
				integration on a 200-mm fully CMOS-compatible silicon photonics platform}},\
	}\href@noop {} {\bibfield  {journal} {\bibinfo  {journal} {IEEE Journal of
				Selected Topics in Quantum Electronics}\ }\textbf {\bibinfo {volume} {25}},\
		\bibinfo {pages} {1} (\bibinfo {year} {2019})}\BibitemShut {NoStop}%
	\bibitem [{\citenamefont {Boitier}\ \emph {et~al.}(2014)\citenamefont
		{Boitier}, \citenamefont {Orieux}, \citenamefont {Autebert}, \citenamefont
		{Lema\^{\i}tre}, \citenamefont {Galopin}, \citenamefont {Manquest},
		\citenamefont {Sirtori}, \citenamefont {Favero}, \citenamefont {Leo},\ and\
		\citenamefont {Ducci}}]{Boitier14}%
	\BibitemOpen
	\bibfield  {author} {\bibinfo {author} {\bibfnamefont {F.}~\bibnamefont
			{Boitier}}, \bibinfo {author} {\bibfnamefont {A.}~\bibnamefont {Orieux}},
		\bibinfo {author} {\bibfnamefont {C.}~\bibnamefont {Autebert}}, \bibinfo
		{author} {\bibfnamefont {A.}~\bibnamefont {Lema\^{\i}tre}}, \bibinfo {author}
		{\bibfnamefont {E.}~\bibnamefont {Galopin}}, \bibinfo {author} {\bibfnamefont
			{C.}~\bibnamefont {Manquest}}, \bibinfo {author} {\bibfnamefont
			{C.}~\bibnamefont {Sirtori}}, \bibinfo {author} {\bibfnamefont
			{I.}~\bibnamefont {Favero}}, \bibinfo {author} {\bibfnamefont
			{G.}~\bibnamefont {Leo}},\ and\ \bibinfo {author} {\bibfnamefont
			{S.}~\bibnamefont {Ducci}},\ }\bibfield  {title} {\bibinfo {title}
		{Electrically injected photon-pair source at room temperature},\ }\href@noop
	{} {\bibfield  {journal} {\bibinfo  {journal} {Phys. Rev. Lett.}\ }\textbf
		{\bibinfo {volume} {112}},\ \bibinfo {pages} {183901} (\bibinfo {year}
		{2014})}\BibitemShut {NoStop}%
	\bibitem [{\citenamefont {Kang}\ \emph {et~al.}(2016)\citenamefont {Kang},
		\citenamefont {Anirban},\ and\ \citenamefont {Helmy}}]{Kang16}%
	\BibitemOpen
	\bibfield  {author} {\bibinfo {author} {\bibfnamefont {D.}~\bibnamefont
			{Kang}}, \bibinfo {author} {\bibfnamefont {A.}~\bibnamefont {Anirban}},\ and\
		\bibinfo {author} {\bibfnamefont {A.~S.}\ \bibnamefont {Helmy}},\ }\bibfield
	{title} {\bibinfo {title} {Monolithic semiconductor chips as a source for
			broadband wavelength-multiplexed polarization entangled photons},\
	}\href@noop {} {\bibfield  {journal} {\bibinfo  {journal} {Optics Express}\
		}\textbf {\bibinfo {volume} {24}},\ \bibinfo {pages} {15160} (\bibinfo {year}
		{2016})}\BibitemShut {NoStop}%
	\bibitem [{\citenamefont {Wengerowsky}\ \emph {et~al.}(2018)\citenamefont
		{Wengerowsky}, \citenamefont {Joshi}, \citenamefont {Steinlechner},
		\citenamefont {H{\"u}bel},\ and\ \citenamefont {Ursin}}]{Wengerowsky18}%
	\BibitemOpen
	\bibfield  {author} {\bibinfo {author} {\bibfnamefont {S.}~\bibnamefont
			{Wengerowsky}}, \bibinfo {author} {\bibfnamefont {S.~K.}\ \bibnamefont
			{Joshi}}, \bibinfo {author} {\bibfnamefont {F.}~\bibnamefont {Steinlechner}},
		\bibinfo {author} {\bibfnamefont {H.}~\bibnamefont {H{\"u}bel}},\ and\
		\bibinfo {author} {\bibfnamefont {R.}~\bibnamefont {Ursin}},\ }\bibfield
	{title} {\bibinfo {title} {An entanglement-based wavelength-multiplexed
			quantum communication network},\ }\href@noop {} {\bibfield  {journal}
		{\bibinfo  {journal} {Nature}\ }\textbf {\bibinfo {volume} {564}},\ \bibinfo
		{pages} {225} (\bibinfo {year} {2018})}\BibitemShut {NoStop}%
	\bibitem [{\citenamefont {Kues}\ \emph {et~al.}(2017)\citenamefont {Kues},
		\citenamefont {Reimer}, \citenamefont {Roztocki}, \citenamefont {Cort{\'e}s},
		\citenamefont {Sciara}, \citenamefont {Wetzel}, \citenamefont {Zhang},
		\citenamefont {Cino}, \citenamefont {Chu}, \citenamefont {Little} \emph
		{et~al.}}]{Kues17}%
	\BibitemOpen
	\bibfield  {author} {\bibinfo {author} {\bibfnamefont {M.}~\bibnamefont
			{Kues}}, \bibinfo {author} {\bibfnamefont {C.}~\bibnamefont {Reimer}},
		\bibinfo {author} {\bibfnamefont {P.}~\bibnamefont {Roztocki}}, \bibinfo
		{author} {\bibfnamefont {L.~R.}\ \bibnamefont {Cort{\'e}s}}, \bibinfo
		{author} {\bibfnamefont {S.}~\bibnamefont {Sciara}}, \bibinfo {author}
		{\bibfnamefont {B.}~\bibnamefont {Wetzel}}, \bibinfo {author} {\bibfnamefont
			{Y.}~\bibnamefont {Zhang}}, \bibinfo {author} {\bibfnamefont
			{A.}~\bibnamefont {Cino}}, \bibinfo {author} {\bibfnamefont {S.~T.}\
			\bibnamefont {Chu}}, \bibinfo {author} {\bibfnamefont {B.~E.}\ \bibnamefont
			{Little}}, \emph {et~al.},\ }\bibfield  {title} {\bibinfo {title} {On-chip
			generation of high-dimensional entangled quantum states and their coherent
			control},\ }\href@noop {} {\bibfield  {journal} {\bibinfo  {journal}
			{Nature}\ }\textbf {\bibinfo {volume} {546}},\ \bibinfo {pages} {622}
		(\bibinfo {year} {2017})}\BibitemShut {NoStop}%
	\bibitem [{\citenamefont {Reimer}\ \emph {et~al.}(2019)\citenamefont {Reimer},
		\citenamefont {Sciara}, \citenamefont {Roztocki}, \citenamefont {Islam},
		\citenamefont {Cort{\'e}s}, \citenamefont {Zhang}, \citenamefont {Fischer},
		\citenamefont {Loranger}, \citenamefont {Kashyap}, \citenamefont {Cino} \emph
		{et~al.}}]{Reimer19}%
	\BibitemOpen
	\bibfield  {author} {\bibinfo {author} {\bibfnamefont {C.}~\bibnamefont
			{Reimer}}, \bibinfo {author} {\bibfnamefont {S.}~\bibnamefont {Sciara}},
		\bibinfo {author} {\bibfnamefont {P.}~\bibnamefont {Roztocki}}, \bibinfo
		{author} {\bibfnamefont {M.}~\bibnamefont {Islam}}, \bibinfo {author}
		{\bibfnamefont {L.~R.}\ \bibnamefont {Cort{\'e}s}}, \bibinfo {author}
		{\bibfnamefont {Y.}~\bibnamefont {Zhang}}, \bibinfo {author} {\bibfnamefont
			{B.}~\bibnamefont {Fischer}}, \bibinfo {author} {\bibfnamefont
			{S.}~\bibnamefont {Loranger}}, \bibinfo {author} {\bibfnamefont
			{R.}~\bibnamefont {Kashyap}}, \bibinfo {author} {\bibfnamefont
			{A.}~\bibnamefont {Cino}}, \emph {et~al.},\ }\bibfield  {title} {\bibinfo
		{title} {High-dimensional one-way quantum processing implemented on d-level
			cluster states},\ }\href@noop {} {\bibfield  {journal} {\bibinfo  {journal}
			{Nature Physics}\ }\textbf {\bibinfo {volume} {15}},\ \bibinfo {pages} {148}
		(\bibinfo {year} {2019})}\BibitemShut {NoStop}%
	\bibitem [{\citenamefont {Imany}\ \emph {et~al.}(2019)\citenamefont {Imany},
		\citenamefont {Jaramillo-Villegas}, \citenamefont {Alshaykh}, \citenamefont
		{Lukens}, \citenamefont {Odele}, \citenamefont {Moore}, \citenamefont
		{Leaird}, \citenamefont {Qi},\ and\ \citenamefont {Weiner}}]{Imany19}%
	\BibitemOpen
	\bibfield  {author} {\bibinfo {author} {\bibfnamefont {P.}~\bibnamefont
			{Imany}}, \bibinfo {author} {\bibfnamefont {J.~A.}\ \bibnamefont
			{Jaramillo-Villegas}}, \bibinfo {author} {\bibfnamefont {M.~S.}\ \bibnamefont
			{Alshaykh}}, \bibinfo {author} {\bibfnamefont {J.~M.}\ \bibnamefont
			{Lukens}}, \bibinfo {author} {\bibfnamefont {O.~D.}\ \bibnamefont {Odele}},
		\bibinfo {author} {\bibfnamefont {A.~J.}\ \bibnamefont {Moore}}, \bibinfo
		{author} {\bibfnamefont {D.~E.}\ \bibnamefont {Leaird}}, \bibinfo {author}
		{\bibfnamefont {M.}~\bibnamefont {Qi}},\ and\ \bibinfo {author}
		{\bibfnamefont {A.~M.}\ \bibnamefont {Weiner}},\ }\bibfield  {title}
	{\bibinfo {title} {High-dimensional optical quantum logic in large
			operational spaces},\ }\href@noop {} {\bibfield  {journal} {\bibinfo
			{journal} {npj Quantum Information}\ }\textbf {\bibinfo {volume} {5}},\
		\bibinfo {pages} {1} (\bibinfo {year} {2019})}\BibitemShut {NoStop}%
	\bibitem [{\citenamefont {Abouraddy}\ \emph {et~al.}(2007)\citenamefont
		{Abouraddy}, \citenamefont {Yarnall}, \citenamefont {Saleh},\ and\
		\citenamefont {Teich}}]{Abouraddy07}%
	\BibitemOpen
	\bibfield  {author} {\bibinfo {author} {\bibfnamefont {A.~F.}\ \bibnamefont
			{Abouraddy}}, \bibinfo {author} {\bibfnamefont {T.}~\bibnamefont {Yarnall}},
		\bibinfo {author} {\bibfnamefont {B.~E.}\ \bibnamefont {Saleh}},\ and\
		\bibinfo {author} {\bibfnamefont {M.~C.}\ \bibnamefont {Teich}},\ }\bibfield
	{title} {\bibinfo {title} {Violation of {Bell’s} inequality with continuous
			spatial variables},\ }\href@noop {} {\bibfield  {journal} {\bibinfo
			{journal} {Phys. Rev. A}\ }\textbf {\bibinfo {volume} {75}},\ \bibinfo
		{pages} {052114} (\bibinfo {year} {2007})}\BibitemShut {NoStop}%
	\bibitem [{\citenamefont {Fabre}\ \emph {et~al.}(2022)\citenamefont {Fabre},
		\citenamefont {Amanti}, \citenamefont {Baboux}, \citenamefont {Keller},
		\citenamefont {Ducci},\ and\ \citenamefont {Milman}}]{Fabre22}%
	\BibitemOpen
	\bibfield  {author} {\bibinfo {author} {\bibfnamefont {N.}~\bibnamefont
			{Fabre}}, \bibinfo {author} {\bibfnamefont {M.}~\bibnamefont {Amanti}},
		\bibinfo {author} {\bibfnamefont {F.}~\bibnamefont {Baboux}}, \bibinfo
		{author} {\bibfnamefont {A.}~\bibnamefont {Keller}}, \bibinfo {author}
		{\bibfnamefont {S.}~\bibnamefont {Ducci}},\ and\ \bibinfo {author}
		{\bibfnamefont {P.}~\bibnamefont {Milman}},\ }\bibfield  {title} {\bibinfo
		{title} {The {Hong--Ou--Mandel} experiment: from photon indistinguishability
			to continuous-variable quantum computing},\ }\href@noop {} {\bibfield
		{journal} {\bibinfo  {journal} {The European Physical Journal D}\ }\textbf
		{\bibinfo {volume} {76}},\ \bibinfo {pages} {196} (\bibinfo {year}
		{2022})}\BibitemShut {NoStop}%
	\bibitem [{\citenamefont {Descamps}\ \emph {et~al.}(2023)\citenamefont
		{Descamps}, \citenamefont {Fabre}, \citenamefont {Keller},\ and\
		\citenamefont {Milman}}]{Descamps23}%
	\BibitemOpen
	\bibfield  {author} {\bibinfo {author} {\bibfnamefont {E.}~\bibnamefont
			{Descamps}}, \bibinfo {author} {\bibfnamefont {N.}~\bibnamefont {Fabre}},
		\bibinfo {author} {\bibfnamefont {A.}~\bibnamefont {Keller}},\ and\ \bibinfo
		{author} {\bibfnamefont {P.}~\bibnamefont {Milman}},\ }\bibfield  {title}
	{\bibinfo {title} {Quantum metrology using time-frequency as quantum
			continuous variables: Resources, sub-shot-noise precision and phase space
			representation},\ }\href@noop {} {\bibfield  {journal} {\bibinfo  {journal}
			{Phys. Rev. Lett.}\ }\textbf {\bibinfo {volume} {131}},\ \bibinfo {pages}
		{030801} (\bibinfo {year} {2023})}\BibitemShut {NoStop}%
	\bibitem [{\citenamefont {Silverstone}\ \emph {et~al.}(2014)\citenamefont
		{Silverstone}, \citenamefont {Bonneau}, \citenamefont {Ohira}, \citenamefont
		{Suzuki}, \citenamefont {Yoshida}, \citenamefont {Iizuka}, \citenamefont
		{Ezaki}, \citenamefont {Natarajan}, \citenamefont {Tanner}, \citenamefont
		{Hadfield} \emph {et~al.}}]{Silverstone14}%
	\BibitemOpen
	\bibfield  {author} {\bibinfo {author} {\bibfnamefont {J.~W.}\ \bibnamefont
			{Silverstone}}, \bibinfo {author} {\bibfnamefont {D.}~\bibnamefont
			{Bonneau}}, \bibinfo {author} {\bibfnamefont {K.}~\bibnamefont {Ohira}},
		\bibinfo {author} {\bibfnamefont {N.}~\bibnamefont {Suzuki}}, \bibinfo
		{author} {\bibfnamefont {H.}~\bibnamefont {Yoshida}}, \bibinfo {author}
		{\bibfnamefont {N.}~\bibnamefont {Iizuka}}, \bibinfo {author} {\bibfnamefont
			{M.}~\bibnamefont {Ezaki}}, \bibinfo {author} {\bibfnamefont {C.~M.}\
			\bibnamefont {Natarajan}}, \bibinfo {author} {\bibfnamefont {M.~G.}\
			\bibnamefont {Tanner}}, \bibinfo {author} {\bibfnamefont {R.~H.}\
			\bibnamefont {Hadfield}}, \emph {et~al.},\ }\bibfield  {title} {\bibinfo
		{title} {On-chip quantum interference between silicon photon-pair sources},\
	}\href@noop {} {\bibfield  {journal} {\bibinfo  {journal} {Nature Photonics}\
		}\textbf {\bibinfo {volume} {8}},\ \bibinfo {pages} {104} (\bibinfo {year}
		{2014})}\BibitemShut {NoStop}%
	\bibitem [{\citenamefont {Francesconi}\ \emph {et~al.}(2022)\citenamefont
		{Francesconi}, \citenamefont {Raymond}, \citenamefont {Duhamel},
		\citenamefont {Filloux}, \citenamefont {Lema{\^\i}tre}, \citenamefont
		{Milman}, \citenamefont {Amanti}, \citenamefont {Baboux},\ and\ \citenamefont
		{Ducci}}]{Francesconi22}%
	\BibitemOpen
	\bibfield  {author} {\bibinfo {author} {\bibfnamefont {S.}~\bibnamefont
			{Francesconi}}, \bibinfo {author} {\bibfnamefont {A.}~\bibnamefont
			{Raymond}}, \bibinfo {author} {\bibfnamefont {R.}~\bibnamefont {Duhamel}},
		\bibinfo {author} {\bibfnamefont {P.}~\bibnamefont {Filloux}}, \bibinfo
		{author} {\bibfnamefont {A.}~\bibnamefont {Lema{\^\i}tre}}, \bibinfo {author}
		{\bibfnamefont {P.}~\bibnamefont {Milman}}, \bibinfo {author} {\bibfnamefont
			{M.~I.}\ \bibnamefont {Amanti}}, \bibinfo {author} {\bibfnamefont
			{F.}~\bibnamefont {Baboux}},\ and\ \bibinfo {author} {\bibfnamefont
			{S.}~\bibnamefont {Ducci}},\ }\bibfield  {title} {\bibinfo {title} {On-chip
			generation of hybrid polarization-frequency entangled biphoton states},\
	}\href@noop {} {\bibfield  {journal} {\bibinfo  {journal} {submitted to
				Photonics Research}\ } (\bibinfo {year} {2022})}\BibitemShut {NoStop}%
	\bibitem [{\citenamefont {Kwiat}(1997)}]{Kwiat97}%
	\BibitemOpen
	\bibfield  {author} {\bibinfo {author} {\bibfnamefont {P.~G.}\ \bibnamefont
			{Kwiat}},\ }\bibfield  {title} {\bibinfo {title} {Hyper-entangled states},\
	}\href@noop {} {\bibfield  {journal} {\bibinfo  {journal} {Journal of Modern
				Optics}\ }\textbf {\bibinfo {volume} {44}},\ \bibinfo {pages} {2173}
		(\bibinfo {year} {1997})}\BibitemShut {NoStop}%
	\bibitem [{\citenamefont {Xie}\ \emph {et~al.}(2015)\citenamefont {Xie},
		\citenamefont {Zhong}, \citenamefont {Shrestha}, \citenamefont {Xu},
		\citenamefont {Liang}, \citenamefont {Gong}, \citenamefont {Bienfang},
		\citenamefont {Restelli}, \citenamefont {Shapiro}, \citenamefont {Wong} \emph
		{et~al.}}]{Xie15}%
	\BibitemOpen
	\bibfield  {author} {\bibinfo {author} {\bibfnamefont {Z.}~\bibnamefont
			{Xie}}, \bibinfo {author} {\bibfnamefont {T.}~\bibnamefont {Zhong}}, \bibinfo
		{author} {\bibfnamefont {S.}~\bibnamefont {Shrestha}}, \bibinfo {author}
		{\bibfnamefont {X.}~\bibnamefont {Xu}}, \bibinfo {author} {\bibfnamefont
			{J.}~\bibnamefont {Liang}}, \bibinfo {author} {\bibfnamefont {Y.-X.}\
			\bibnamefont {Gong}}, \bibinfo {author} {\bibfnamefont {J.~C.}\ \bibnamefont
			{Bienfang}}, \bibinfo {author} {\bibfnamefont {A.}~\bibnamefont {Restelli}},
		\bibinfo {author} {\bibfnamefont {J.~H.}\ \bibnamefont {Shapiro}}, \bibinfo
		{author} {\bibfnamefont {F.~N.}\ \bibnamefont {Wong}}, \emph {et~al.},\
	}\bibfield  {title} {\bibinfo {title} {Harnessing high-dimensional
			hyperentanglement through a biphoton frequency comb},\ }\href@noop {}
	{\bibfield  {journal} {\bibinfo  {journal} {Nature Photonics}\ }\textbf
		{\bibinfo {volume} {9}},\ \bibinfo {pages} {536} (\bibinfo {year}
		{2015})}\BibitemShut {NoStop}%
	\bibitem [{\citenamefont {Steinlechner}\ \emph {et~al.}(2017)\citenamefont
		{Steinlechner}, \citenamefont {Ecker}, \citenamefont {Fink}, \citenamefont
		{Liu}, \citenamefont {Bavaresco}, \citenamefont {Huber}, \citenamefont
		{Scheidl},\ and\ \citenamefont {Ursin}}]{Steinlechner17}%
	\BibitemOpen
	\bibfield  {author} {\bibinfo {author} {\bibfnamefont {F.}~\bibnamefont
			{Steinlechner}}, \bibinfo {author} {\bibfnamefont {S.}~\bibnamefont {Ecker}},
		\bibinfo {author} {\bibfnamefont {M.}~\bibnamefont {Fink}}, \bibinfo {author}
		{\bibfnamefont {B.}~\bibnamefont {Liu}}, \bibinfo {author} {\bibfnamefont
			{J.}~\bibnamefont {Bavaresco}}, \bibinfo {author} {\bibfnamefont
			{M.}~\bibnamefont {Huber}}, \bibinfo {author} {\bibfnamefont
			{T.}~\bibnamefont {Scheidl}},\ and\ \bibinfo {author} {\bibfnamefont
			{R.}~\bibnamefont {Ursin}},\ }\bibfield  {title} {\bibinfo {title}
		{Distribution of high-dimensional entanglement via an intra-city free-space
			link},\ }\href@noop {} {\bibfield  {journal} {\bibinfo  {journal} {Nature
				Communications}\ }\textbf {\bibinfo {volume} {8}},\ \bibinfo {pages} {1}
		(\bibinfo {year} {2017})}\BibitemShut {NoStop}%
	\bibitem [{\citenamefont {Kim}\ \emph {et~al.}(2021)\citenamefont {Kim},
		\citenamefont {Kim}, \citenamefont {Im}, \citenamefont {Lee}, \citenamefont
		{Chae}, \citenamefont {Scarcelli},\ and\ \citenamefont {Kim}}]{Kim21}%
	\BibitemOpen
	\bibfield  {author} {\bibinfo {author} {\bibfnamefont {J.-H.}\ \bibnamefont
			{Kim}}, \bibinfo {author} {\bibfnamefont {Y.}~\bibnamefont {Kim}}, \bibinfo
		{author} {\bibfnamefont {D.-G.}\ \bibnamefont {Im}}, \bibinfo {author}
		{\bibfnamefont {C.-H.}\ \bibnamefont {Lee}}, \bibinfo {author} {\bibfnamefont
			{J.-W.}\ \bibnamefont {Chae}}, \bibinfo {author} {\bibfnamefont
			{G.}~\bibnamefont {Scarcelli}},\ and\ \bibinfo {author} {\bibfnamefont
			{Y.-H.}\ \bibnamefont {Kim}},\ }\bibfield  {title} {\bibinfo {title}
		{Noise-resistant quantum communications using hyperentanglement},\
	}\href@noop {} {\bibfield  {journal} {\bibinfo  {journal} {Optica}\ }\textbf
		{\bibinfo {volume} {8}},\ \bibinfo {pages} {1524} (\bibinfo {year}
		{2021})}\BibitemShut {NoStop}%
	\bibitem [{\citenamefont {Coldren}\ \emph {et~al.}(2012)\citenamefont
		{Coldren}, \citenamefont {Corzine},\ and\ \citenamefont
		{Mashanovitch}}]{Coupled_Mode_Theory}%
	\BibitemOpen
	\bibfield  {author} {\bibinfo {author} {\bibfnamefont {L.~A.}\ \bibnamefont
			{Coldren}}, \bibinfo {author} {\bibfnamefont {S.~W.}\ \bibnamefont
			{Corzine}},\ and\ \bibinfo {author} {\bibfnamefont {M.}~\bibnamefont
			{Mashanovitch}},\ }\bibinfo {title} {Diode lasers and photonic integrated
		circuits}\ (\bibinfo  {publisher} {N.J: Wiley},\ \bibinfo {year} {2012})\
	\bibinfo {edition} {2nd}\ ed.\BibitemShut {Stop}%
	\bibitem [{\citenamefont {De~Rossi}\ \emph {et~al.}(2005)\citenamefont
		{De~Rossi}, \citenamefont {Ortiz}, \citenamefont {Calligaro}, \citenamefont
		{Lanco}, \citenamefont {Ducci}, \citenamefont {Berger},\ and\ \citenamefont
		{Sagnes}}]{FP}%
	\BibitemOpen
	\bibfield  {author} {\bibinfo {author} {\bibfnamefont {A.}~\bibnamefont
			{De~Rossi}}, \bibinfo {author} {\bibfnamefont {V.}~\bibnamefont {Ortiz}},
		\bibinfo {author} {\bibfnamefont {M.}~\bibnamefont {Calligaro}}, \bibinfo
		{author} {\bibfnamefont {L.}~\bibnamefont {Lanco}}, \bibinfo {author}
		{\bibfnamefont {S.}~\bibnamefont {Ducci}}, \bibinfo {author} {\bibfnamefont
			{V.}~\bibnamefont {Berger}},\ and\ \bibinfo {author} {\bibfnamefont
			{I.}~\bibnamefont {Sagnes}},\ }\bibfield  {title} {\bibinfo {title}
		{{Measuring propagation loss in a multimode semiconductor waveguide}},\
	}\href@noop {} {\bibfield  {journal} {\bibinfo  {journal} {Journal of Applied
				Physics}\ }\textbf {\bibinfo {volume} {97}},\ \bibinfo {pages} {073105}
		(\bibinfo {year} {2005})}\BibitemShut {NoStop}%
	\bibitem [{\citenamefont {Kaiser}\ \emph {et~al.}(2014)\citenamefont {Kaiser},
		\citenamefont {Ngah}, \citenamefont {Issautier}, \citenamefont {Delord},
		\citenamefont {Aktas}, \citenamefont {D׳Auria}, \citenamefont {{De
				Micheli}}, \citenamefont {Kastberg}, \citenamefont {Labonté}, \citenamefont
		{Alibart}, \citenamefont {Martin},\ and\ \citenamefont
		{Tanzilli}}]{KAISER20147}%
	\BibitemOpen
	\bibfield  {author} {\bibinfo {author} {\bibfnamefont {F.}~\bibnamefont
			{Kaiser}}, \bibinfo {author} {\bibfnamefont {L.}~\bibnamefont {Ngah}},
		\bibinfo {author} {\bibfnamefont {A.}~\bibnamefont {Issautier}}, \bibinfo
		{author} {\bibfnamefont {T.}~\bibnamefont {Delord}}, \bibinfo {author}
		{\bibfnamefont {D.}~\bibnamefont {Aktas}}, \bibinfo {author} {\bibfnamefont
			{V.}~\bibnamefont {D'Auria}}, \bibinfo {author} {\bibfnamefont
			{M.}~\bibnamefont {{De Micheli}}}, \bibinfo {author} {\bibfnamefont
			{A.}~\bibnamefont {Kastberg}}, \bibinfo {author} {\bibfnamefont
			{L.}~\bibnamefont {Labonté}}, \bibinfo {author} {\bibfnamefont
			{O.}~\bibnamefont {Alibart}}, \bibinfo {author} {\bibfnamefont
			{A.}~\bibnamefont {Martin}},\ and\ \bibinfo {author} {\bibfnamefont
			{S.}~\bibnamefont {Tanzilli}},\ }\bibfield  {title} {\bibinfo {title}
		{Polarization entangled photon-pair source based on quantum nonlinear
			photonics and interferometry},\ }\href@noop {} {\bibfield  {journal}
		{\bibinfo  {journal} {Optics Communications}\ }\textbf {\bibinfo {volume}
			{327}},\ \bibinfo {pages} {7} (\bibinfo {year} {2014})}\BibitemShut {NoStop}%
	\bibitem [{\citenamefont {Franson}(1991)}]{Franson_vis}%
	\BibitemOpen
	\bibfield  {author} {\bibinfo {author} {\bibfnamefont {J.~D.}\ \bibnamefont
			{Franson}},\ }\bibfield  {title} {\bibinfo {title} {Violations of a simple
			inequality for classical fields},\ }\href@noop {} {\bibfield  {journal}
		{\bibinfo  {journal} {Physical Review Letters}\ }\textbf {\bibinfo {volume}
			{67}},\ \bibinfo {pages} {290} (\bibinfo {year} {1991})}\BibitemShut
	{NoStop}%
\end{thebibliography}


\end{document}